\documentclass{aa}
\usepackage{epsfig}
\usepackage{rotating}
\usepackage{txfonts}
\usepackage{longtable,lscape}
\newcommand{\gesssim}{\mathrel{\hbox{\rlap{\hbox{\lower4pt\hbox{$\sim$}}}\hbox{$>$}}}}

\def\hii{H~{\sc ii}}

\begin{document}

\title{The evolution of the Galactic metallicity gradient \\ 
from high-resolution spectroscopy of open clusters}

\subtitle{}

\author{Laura Magrini, Paola Sestito, Sofia Randich, 
Daniele Galli}

\authorrunning{Magrini et al.}

\offprints{L. Magrini}

\institute{INAF-Osservatorio Astrofisico di Arcetri, Largo E.~Fermi 5,
             I-50125 Firenze, Italy\\
\email{laura, sestito, randich, galli@arcetri.astro.it}
}

\titlerunning{Evolution of the Galactic metallicity gradient}
\date{Received Date: Accepted Date}

\abstract
{Open clusters offer a unique possibility to study the time evolution
of the radial metallicity  gradients of several elements in our Galaxy,
because they span large intervals in age and Galactocentric distance,
and both quantities can be more accurately derived than for field stars.}
{We re-address the issue of the Galactic metallicity gradient and its
time evolution by comparing the empirical gradients traced by
a sample of 45 open clusters
with a chemical evolution model of the Galaxy.}
{At variance with previous similar studies, we have collected 
from the literature only abundances derived
from high--resolution spectra. The clusters have
Galactocentric distances $7 \lesssim R_{\rm GC} \lesssim 22$~kpc and
ages from $\sim 30$~Myr to 11~Gyr. We also
consider
the $\alpha$-elements Si, Ca, Ti, and the iron-peak elements Cr and
Ni. Cepheids trace instead the present-day Fe gradient in the inner
parts of the disk.}
{The data for iron-peak and $\alpha$-elements indicate a steep
metallicity gradient for $R_{\rm GC}\lesssim 12$~kpc and a plateau 
at larger radii. The time evolution of the metallicity distribution is
characterized by a uniform increase of the metallicity at all radii,
preserving the shape of the gradient, with  marginal evidence for a
flattening of the gradient with time in the radial range 7-12~kpc.  
Our model is able to reproduce the main features of the 
metallicity  gradient
and its evolution with an infall law exponentially decreasing with 
radius and with a 
collapse time scale of the order of 8~Gyr at the solar radius. 
This results in a rapid collapse  in the inner regions, 
i.e. $R_{\rm GC}\lesssim 12$~kpc (that we associate with an early
phase of disk formation from the collapse of the halo) and in a slow inflow 
of material per unit area in the outer regions at a constant rate with time (that we associate with accretion from the
intergalactic medium).
An additional uniform inflow per unit disk area would help to better reproduce the metallicity 
plateau at large Galactocentric radii, but it is difficult to reconcile with the present-day radial behaviour of the 
star formation rate.}
{Our results favour a scenario where the Galactic disk is formed inside-out
by the rapid collapse of the halo and by a subsequent continuous 
accretion of intergalactic gas}

\maketitle

\section{Introduction}
\label{intro}

The radial metallicity gradient and its time evolution provide 
strong constraints on our understanding of the formation and the
evolution of galaxies.  Detailed models of Galactic chemical evolution
(GCE) have been continuously improved over the last decades (e.g., Lacey
\& Fall \cite{lacey}; Tosi \cite{tosi88}; Ferrini et al.~\cite{ferrini92},
\cite{ferrini94}; Giovagnoli \& Tosi \cite{giova}; Chiappini, Matteucci
\& Gratton \cite{chiappini97}; Boissier \& Prantzos \cite{boissier99};
Hou et al.~\cite{hou00}; Chiappini et al.~\cite{chiappini01}). However,
the issue of the time evolution of the radial abundance gradients
is far from being settled unequivocally, both from a theoretical (e.g.,
Goetz \& Koeppen~\cite{goetz92}; Koeppen~\cite{koeppen94}; Moll{\`a}
et al.~\cite{molla97}; Henry \& Worthey~\cite{henry99}; Chiappini et
al.~\cite{chiappini01}) and an observational (e.g., Friel \cite{friel95};
Friel et al.~\cite{friel02}; Maciel~\cite{maciel00}; Maciel et
al.~\cite{maciel03}; Stanghellini et al.~\cite{stanghellini06}) point
of view.  While most GCE models are able to reproduce the present-day
radial distribution of several chemical elements derived from sample
objects representative of the present-day composition of the interstellar
medium (ISM), such as, e.g., \hii\, regions, Cepheids, super-giant stars,
they generally disagree on the predicted behaviour of its time evolution.

In this respect, GCE models can be divided in two groups: models
which predict a {\em steepening} of the metallicity gradient with
time (e.g., Tosi \cite{tosi88}; Chiappini et al.~\cite{chiappini97};
Chiappini et al.~\cite{chiappini01}) and those which predict a {\em
flattening} with time (e.g., Moll\`a et al.~\cite{molla97}; Portinari \&
Chiosi~\cite{portinari99}; Hou et al.~\cite{hou00}).  The main differences
between
the two groups of models are the efficiency of the enrichment processes in
the inner andf outer regions of the Galactic disk (the star formation
rate, SFR), and the nature of the material (primordial or pre-enriched)
falling from the halo onto the disk (the infall). 

In models of the former group the outer disk is pre-enriched by the
previous evolution of the halo, and during the first Gyr its metallicity
is affected very little by the relatively lower SFR in the disk. Since
the infall decreases rapidly with Galactocentric radius, reflecting the
strongly concentrated mass distribution in the halo, a larger amount
of metal poor material falls in the central regions rather than in the
outer Galactic disk. Thus, a positive metallicity gradient is initially
established. When the halo collapse phase is over, the metallicity
increases more rapidly in the inner disk than in the outer Galaxy,
because of a combination of higher SFR at small radii and continuous
dilution by infall at large radii, reversing the sign of the gradient.

In contrast, in models of the latter group, the initially vigorous star
formation activity in the inner disk and the metal-poor composition of
the infalling gas produce a rapid increase of the metallicity near the
Galactic center. In the inner regions the metallicity increases rapidly,
reaching its final value in the first 2--3 Gyr of disk evolution, whereas
the enrichment of the outer disk is slower, resulting in a progressive
flattening of the radial gradient.  In this group of models, various
infall rates have been assumed: a very rapid one simulating the formation
of the disk from a monolithic collapse of the halo, or one much diluted
with time, simulating a hierarchical formation of the disk by continuous
infall of gas from the intergalactic medium (IGM). These rates produce
different time evolution of the metallicity gradient: a rapid flattening
or a uniform increase of the metallicity with essentially no change of
slope, respectively (e.g., Moll\`a et al.\cite{molla96}).

In principle, it should be possible to discriminate between the
different evolutionary scenario, by comparing the metallicity gradient
of the ``old'' stellar population  e.g., red giant branch (RGB)
stars, planetary nebulae (PNe), and old open clusters (OCs), with the
present-day gradient outlined, e.g., by \hii\,  regions, young stars
(including Cepheids and giant stars) and young OCs.  In external galaxies,
the comparison of PN and \hii\, region abundances, investigated with
similar observational and analysis techniques, has given encouraging
results (e.g., Magrini et al.~\cite{magrini07} for the spiral galaxy M33),
since PNe can be assumed at the same distance as the host galaxy.

In our Galaxy, however, the huge uncertainty on the distances to
Galactic PNe restrict their use as reliable tracers of the past
evolution of metallicity gradients.  On the contrary, the determination
of metallicities of OCs represents a reliable approach to the study of
the time evolution of the metallicity gradient. OCs represent, in fact,
a unique opportunity to study abundances of a large number of elements
in stars of known ages and located at a wide range of well-determined
Galactocentric distances. Thus, they can fully address the issue of the
time evolution of the metallicity gradient, provided that a homogeneous
analysis is performed.

Several spectroscopic studies of chemical abundances in OCs have
been carried out during the last years, both at low- (e.g., Friel
\cite{friel95}; Friel et al.~\cite{friel02}) and high-resolution (e.g.,
Carretta et al.~\cite{carretta04}; Carraro et al.~\cite{carraro04},
\cite{carraro06}, \cite{carraro07}; Yong, Carney \& de Almeida
\cite{yong05}; Sestito et al~\cite{sestito06}, \cite{sestito08}; Bragaglia
et al.~\cite{bragaglia07}). Different authors find different results,
depending on the quality of the data, and in particular on the spectral
resolution, and on the method of analysis adopted. These discrepancies
especially affect metallicity gradients. Also, abundances for the farthest
clusters, which are critical to determine the gradient up to very large
Galactocentric distances, were not available until very recently. For
example, Friel (\cite{friel95}; \cite{friel06}), Carraro, Ng \&
Portinari (\cite{carraro98}) and Friel et al.~(\cite{friel02}) suggest
the presence of a negative [Fe/H] gradient, while Twarog, Ashman \&
Anthony-Twarog (\cite{twarog}) and Corder \& Twarog (\cite{corder})
favor a step-like distribution of the Fe content with Galactocentric
distance. Furthermore, recent results (e.g., Yong et al. \cite{yong05},
Carraro et al.~\cite{carraro07}; Sestito et al.~\cite{sestito07}) suggest
that the gradient becomes flat for Galactocentric radii $R_{\rm GC}$
larger than $\sim 11-12$~kpc.

For the time evolution, evidence for variation so far  has been 
not conclusive.  The majority of studies seem to favor a slight
evolution from steeper gradients in the past to shallower one at the
present time. For example, Friel et al.~(\cite{friel02}), based on
metallicities derived from low resolution spectra, find a flattening of
the {\it overall} gradient from $-0.096$~dex\,kpc$^{-1}$ for ages greater
than 6~Gyr, to $-0.023$~dex\,kpc$^{-1}$ for clusters younger than 2~Gyr;
similarly, Chen et al.~(\cite{chen03}) derive a shallower slope for clusters
younger than 0.8~Gyr than for older ones. On the contrary Salaris et al.
(\cite{salaris04}) find a steeper slope for younger clusters.  
These discrepant results are likely due to: {\em i)} use of different
and/or inhomogeneous abundance datasets;
{\em ii)} the different radial regions of our Galaxy which are analyzed.

In this paper we re-address the issue of the radial metallicity gradient
in our Galaxy observationally and theoretically, exploiting the availability
of abundances based on high resolution spectra for several open
clusters. Specifically, 
we collected and analyzed the most recent metallicity determinations of a large
sample of OCs, considering {\em only measurements based on high resolution
spectroscopy}, which by far are more reliable than those based on photometry
or low resolution spectroscopy. These observations, described in 
Sect.~\ref{sect_data},
are compared with the results of our chemical evolution model, able to
reproduce the main observational features of our Galaxy.  In addition to
Fe, the most recent studies (e.g., Friel et al.~\cite{friel03}; Carretta
et al.~\cite{carretta04}; Friel, Jacobson, \& Pilachowski \cite{friel05};
Yong et al. \cite{yong05}; Sestito et al.~\cite{sestito06}; Randich
et al.~\cite{randich06}; Bragaglia et al.~\cite{bragaglia07}; Sestito
et al.~\cite{sestito07}, \cite{sestito08}) also include the analysis
of other elements. These studies allow us to investigate for the first
time the radial gradient of Fe-peak and $\alpha$-elements based on the
comparison of OC abundances, rather than on field stars, and model
predictions. Finally, we consider the empirical gradient traced by
Cepheids for the inner parts of the disk (R$_{\rm GC} \leq 7$~kpc).

The rest of the paper is organized as follows: in Sect.~\ref{sect_model}
we describe the model; in Sect.~\ref{sect_confronto} we analyze the
time evolution of the radial gradient of several chemical elements,
including Fe and several $\alpha$ and iron-peak elements; 
the abundance ratio radial behaviours are discussed in Sect.~\ref{sect_other}, while 
the results of this study are discussed in Sect.~\ref{sect_discussion}; 
in Sect.~\ref{sect_summary} we summarize our conclusions.
In the Appendix we describe the details of the Galactic chemical evolution model.

\section{The data sample} 
\label{sect_data}

As discussed in Section~1, since 
the early study by Janes (\cite{janes79}) several works have
focused on the determination of the radial metallicity gradient based on OCs.
The first remarkable studies addressing the time evolution of the
gradient through OC data were performed  by Friel
et al.~(\cite{friel02}) and by Chen et al.~(\cite{chen03}). The former
analyzed a sample of low- and moderate-resolution 
data of Galactic OCs over a range in Galactocentric radius from 7 to 16~kpc. 
The latter was based on a collection of metallicities for a very large
sample of OCs up to a distance $\leq$16~kpc from the Galactic center,
whose metallicities and ages had been derived
by different authors and with different methods. 
Therefore, whereas the catalog of Chen et al.~(\cite{chen03}) represents
the largest sample of OCs so far collected to study the gradient,
it is greatly inhomogeneous.

In the last few years, and after the two studies mentioned above,
thanks to state-of-the-art multiplex facilities on 8m class, 
the number of OC with secure metallicity determinations from
high-resolution spectroscopy has greatly increased. Also,
and most important, it has been possible
to extend metallicity determinations
to OCs in the outer disk, beyond 15-16~kpc from the Galactic center.
Finally, abundances of other elements besides iron have been determined.

In this context, the strength of our data sample
is that it combines the largest available sample of high-resolution
observations of OCs, including OCs with Galactocentric distances
between $\sim 7$ and 22~kpc, together with a homogeneous treatment of
their age derivation, attaining larger Galactocentric distances
than in previous samples.

Our dataset of OCs (see Table~\ref{tab_ocs}) is a collection of samples
analyzed by different authors in the literature. Table~\ref{tab_ocs}
lists in Cols.~1 and 2 the sample clusters and their Galactocentric
distances $R_{\rm GC}$, taken from Friel et al.~(\cite{friel02}), when
available, otherwise from Friel (\cite{friel95}, \cite{friel06}); in the
few cases when R$_{\rm GC}$ from Friel was not available, we calculated
this quantity using the cluster distance given in the papers quoted in
the table and assuming a solar Galactocentric distance of 8.5~kpc,
consistent with Friel et al.

In Col.~3 we list the ages of the sample clusters. Since they are
retrieved from different sources, a unique and/or consistent age
determination for  them is not available. As is well known, age is
 very sensitive to the method employed, to the adopted stellar
models in the case of isochrone fitting, and it can vary significantly
from one study to another of the same cluster. In order to obtain a
homogeneous set of ages for the oldest clusters (ages $\geq 0.5$~Gyr),
we decided to use the morphological age indicator $\delta$V (Phelps et
al.~\cite{phelps94}) and the metallicity-dependent calibration of Salaris
et al.~(\cite{salaris04}). For most of the sample clusters $\delta$V
is available from the literature; when not available, we obtained
this quantity from available color-magnitude diagrams. Alternatively,
we determined $\delta$(B-V) and derived $\delta$V from it (see again
Phelps et al.). Note that for the calibrating clusters of Salaris et
al.~(\cite{salaris04}) we directly used their age determination. For
the younger clusters (the first seven lines of Table~\ref{tab_ocs}),
we used the most recent age determinations available in the literature,
such as those obtained with the lithium depletion boundary method. Use
of non-homogeneous age determinations for young clusters will not affect
our results and discussion, given our classification of the OCs in three
major age bins (see below).

Information on the observations and the analysis techniques is listed
in Cols.~5, 6, 7, 8: specifically, we give the signal to noise ratio
(S/N), the resolving power (R), the number and type of stars analyzed
(giant or dwarf stars), and the method of analysis. The values of
[Fe/H] and their errors, references to the observations and abundance
analysis are listed in Cols.~9 and 10. As the table shows, the values of
R and S/N cover a rather wide interval, although for most samples the
resolution is R$\sim$30,000--40,000.  The metallicities are more often
derived through equivalent width measurements, although in a few
cases spectral synthesis was employed.  The analysis is based on giant
stars for most of the intermediate age and old clusters, while for the
young, more close-by cluster main sequence stars have been used. As
discussed by Randich et al.~(\cite{randich06}) for M67 and Pasquini et
al. (\cite{pasquini}) for IC4651, this should not introduce any major
bias for the elements considered here.  The numbers of stars used in
each analysis range from 1 to 15, with the exception of the very rich
set by Paulson et al.~(\cite{paulson03}) for the Hyades (55 dwarf stars).

In order to check whether (and to what extent) use of different abundance
scales might introduce spurious trends or increase the dispersion,
we compared different abundance measurements for the same cluster.
Considering only the most recent studies based on high resolution,
we found 11 cases with more than one [Fe/H] determination available.
The difference between the [Fe/H] values does not depend on  [Fe/H] itself
and the average
is rather small; namely, $\Delta$[Fe/H]=0.12~dex
with a standard deviation of 0.08~dex. This value is somewhat larger (a
factor of $\sim 2$) than the typical error in [Fe/H] of most clusters;
however, we believe that it is  small enough not to affect our results in
a significant way. For  other element abundances, very few clusters have
more than one determination; these studies indicate that the differences
in [X/H] are likely larger and can be as large as 0.2~dex and might
introduce a certain amount of scatter in the observed distributions.

\section{The chemical evolution model}
\label{sect_model}
 
The model adopted in this work is a generalization of the multi-phase
model by Ferrini et al.~(\cite{ferrini92}), originally built for the
solar neighborhood, and subsequently extended to the entire Galaxy
(Ferrini et al.~\cite{ferrini94}), and to other disk galaxies
(e.g., Moll\`a et al.~\cite{molla96}, \cite{molla97}; Moll\`a \&
Diaz~\cite{molla05}; Magrini et al.~\cite{magrini07}).  
We refer
to Ferrini et al.~(\cite{ferrini92},\cite{ferrini94}), Moll\'a
et al.~(\cite{molla96}), and to Magrini et
al.~(\cite{magrini07}) for a detailed description of the general model.
The main assumptions of our model are that the Galaxy disk is formed 
by infall of gas from the halo and from the interstellar medium. 
The adopted infall has an exponentially decreasing law. This produces 
an inside-out formation scenario where the inner parts of the disk evolve more rapidly
than the  outer ones. We consider also an alternative parametrization of the infall 
including a constant amount of gas per unit area in addition to the exponentially decreasing law. 
This latter parametrization is useful in reproducing the metallicity gradient, 
but resulted in an  enhanced star formation rate in the outer regions.      
 
In the Appendix the overall content of our model
and in particular the specific assumptions and parameterizations we
used to reproduce the Galactic disk properties are described in detail.

\section{Radial gradients: comparison between models and observations}
\label{sect_confronto}
 
In Figs.~\ref{iron1}, \ref{fepeak}, \ref{alpha1}, and
\ref{alpha2}, we compare abundances in OCs with
the predictions of the model described in the Appendix  for
the $\alpha$-elements Si, Ca, Ti, and the iron-peak elements, Fe, Ni,
and Cr\footnote{For the classification of elements in the $\alpha$- and
iron-peak we have followed Reddy et al.~(\cite{reddy06}).}.  In general,
the observed abundance distributions show a two-slope feature with a 
change from a steep gradient for  $R_{\rm GC}\lesssim 11-12$~kpc 
and a flat plateau at greater distances, at least in the two 
older age bins. 
For simplicity, we will assume 12~kpc as the threshold 
Galactocentric distance to distinguish the inner 
regions from the outer plateau. We caution that this value
is rather arbitrary since the samples 
in each age bin are limited and there is
a lack of OC data in the radial range 
12--15~kpc which would allow us to better constrain this value. 
We note however that our conclusions would not change by assuming
a smaller (e.g., 11~kpc) or larger (e.g., 13~kpc) radius.
For the Cepheids a change of slope
also is seen  at 7~kpc, with a steepening of the slope for $R_{\rm GC} <
7$~kpc. 

In the following we will discuss separately the iron and other element
gradients.  For a better comparison with the chemical model, we have
divided the cluster ages in three bins, namely ({\em i}\/) age less
than or equal to 0.8~Gyr, ({\em ii}\/) 0.8~Gyr$<$age$\leq 4$~Gyr, and
({\em iii}\/) 4~Gyr$<$age$\leq$ 11~Gyr.

\subsection{Iron}
\label{iron}

\begin{figure} 
\psfig{figure=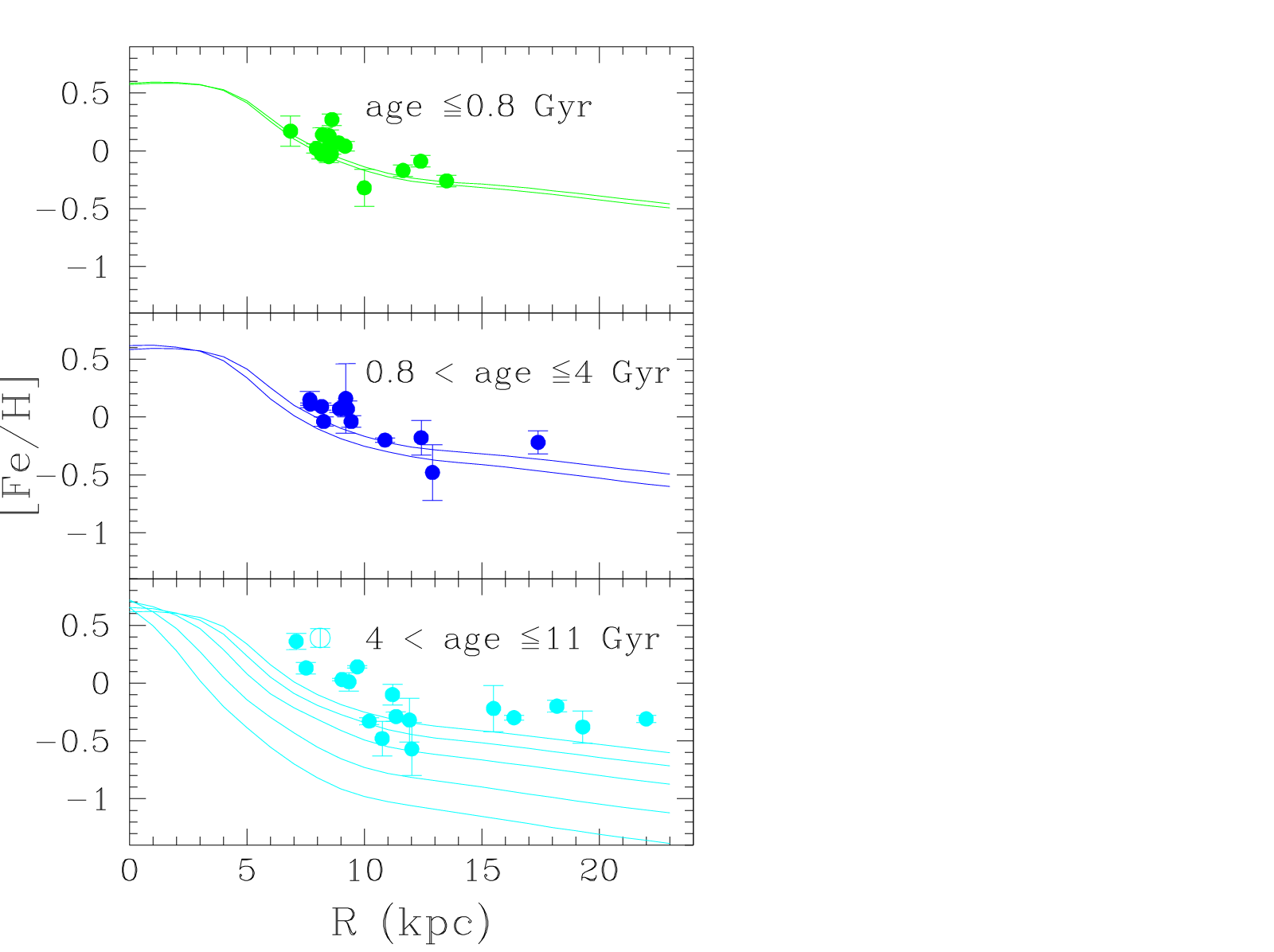, width=12cm, angle=-90} 
\caption{Gradient of [Fe/H]: comparison between high-resolution data of
OCs with the model decribed in the Appendix. Different panels
show abundances for different age bins: top (panel 1), ages $\leq$0.8~Gyr;
middle (panel 2), ages $0.8~\mbox{Gyr} <\mbox{age}\leq 4~\mbox{Gyr}$;
bottom (panel 3), ages $4~\mbox{Gyr} <\mbox{age}\leq 11~\mbox{Gyr}$.  The
models adopted to compare with observations are for 0 and 1 Gyr ago (panel
1), 1 and 4 Gyr ago (panel 2), and 4, 6, 8, 10, 11 Gyr ago (panel 3).}
\label{iron1} 
\end{figure}

\begin{figure}
\psfig{figure=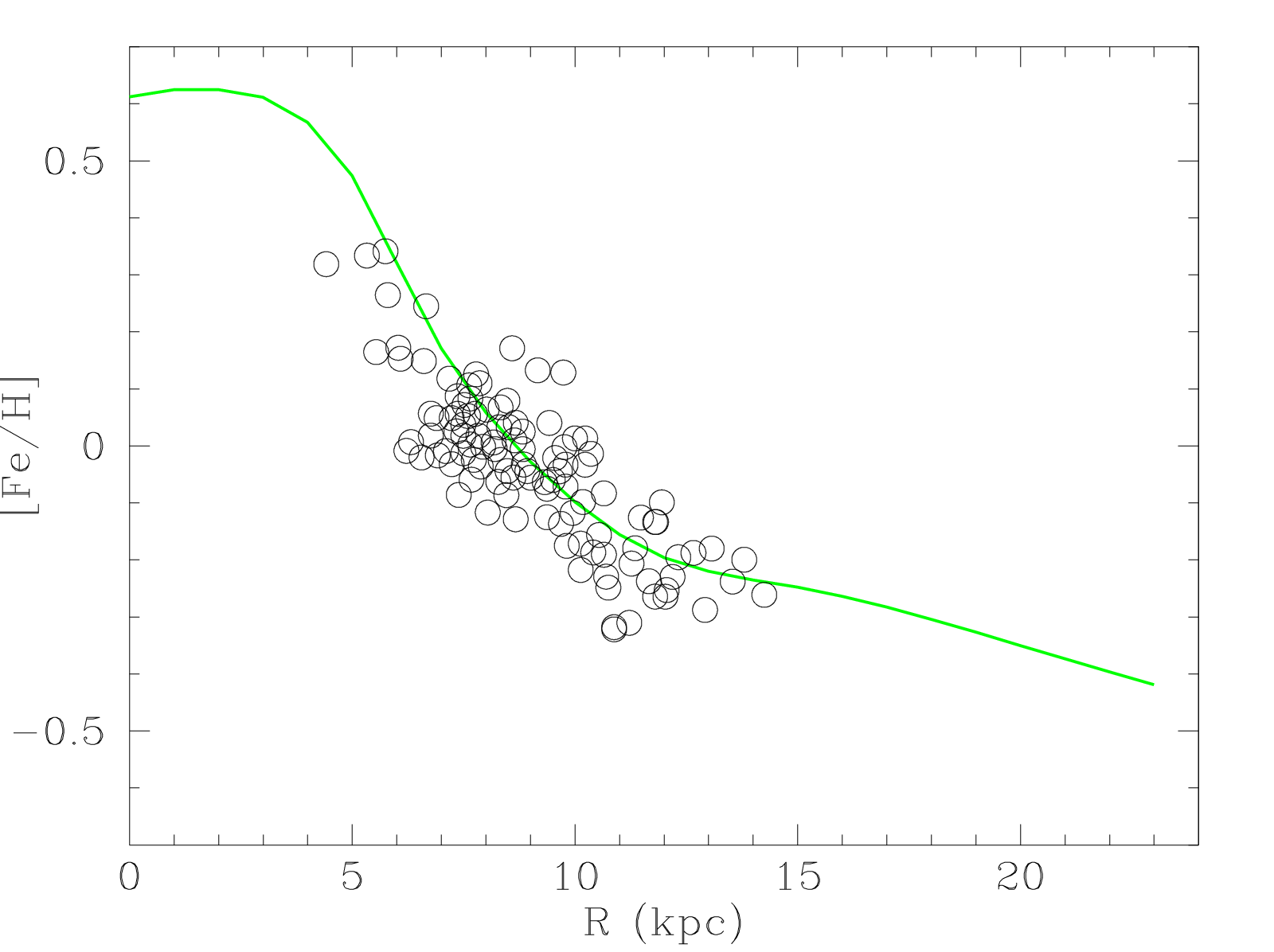, width=7cm, angle=-90} 
\caption{Gradient of [Fe/H]: comparison between Cepheid data (Andrievsky et
al.~\cite{andri04}) and our current model.}
\label{iron2} 
\end{figure}

In Fig.~\ref{iron1} and \ref{iron2} we show the distribution of [Fe/H]
of OCs and Cepheids as a function of the Galactocentric distance
$R_{\rm GC}$, and compare it with
the [Fe/H] behaviour predicted by our model.  For the Fe abundances
of the Cepheids we have used the gradient derived by Andrievsky et
al.~(\cite{andri04}) and Lemasle et al.~(\cite{lemasle07}). We
recall that the Cepheids represent a young population; thus
their data can be compared only with the current model gradient,
while OC data allow a comparison with the model predictions at different
epochs.  

The shape of the Fe gradient of OCs,  at least in the most populated age bin (age between 4 and 11 Gyr)  is
characterized by a negative slope at short radii and a change of slope
at around $\sim$11-12~kpc, resulting in a plateau at larger distances. 
This
behaviour is qualitatively similar to that of Cepheid data, although
the latter cover an inner part of the disk and show a change of slope
around 7~kpc, but do not extend beyond 15~kpc.

More specifically, it is useful to analyze in detail the gradient in the
three radial zones selected, namely $4~\mbox{kpc}\lesssim R_{\rm{GC}}
\lesssim 7~\mbox{kpc}$ (region~I), $7~\mbox{kpc}\lesssim R_{\rm{GC}}
\lesssim 12~\mbox{kpc}$ (region~II), $12~\mbox{Kpc}\lesssim R_{\rm{GC}}
\lesssim 22~\mbox{kpc}$ (region~III). From Table~\ref{tab_grad_iron}
and Fig.~\ref{iron2} it is evident that the current gradient
outlined by the metallicity of Cepheids is very steep in region~I,
where, unfortunately, no high-resolution OC data are available. In
region~II, the gradient traced by Cepheids ($-0.051$ and $-0.044$ dex
kpc$^{-1}$ according to Lemasle et al.~\cite{lemasle07} and Andrievsky et
al.~\cite{andri04}, respectively) is consistent with that outlined
by the youngest OCs ($-0.053$ dex kpc$^{-1}$ --see Sect.~\ref{sect_discussion}).
Finally, as already suggested by the most recent papers (e.g.  Yong et
al.~\cite{yong05}; Carraro et al.~\cite{carraro04}, \cite{carraro07};
Sestito et al.~\cite{sestito08}), data from of OCs in region~III indicate
a very flat outer gradient, with a slope consistent with zero at all
epochs (see also Sect.~\ref{sect_discussion}).

As noted by Andrievsky et al.~(\cite{andri04}), the existence of a
discontinuity in the metallicity gradient at $\sim$11--12~kpc,
observed for both Cepheid and OCs data, is  predicted and
explained by two chemical evolution models. One of them is the model
proposed by Andrievsky et al.~(\cite{andri04}) themselves, and
the other is  the model C presented by Chiappini et
al.~(\cite{chiappini01}).  In particular,
Andrievsky et al.~(\cite{andri04}) assume that the SFR is a
combination of two components: one proportional to the gas surface
density, and the other depending upon the relative velocity of the
interstellar gas and spiral arms.  However, their model predicts a
strong decrease in metallicity beyond $\sim 14$~kpc, whereas the
observed gradient from OCs is flat up to $\sim 22$~kpc.  
On the
other hand, Chiappini et al.~(2001) assume two main accretion
episodes in the lifetime of the Galaxy, the first one forming the
halo and bulge and the second one forming the thin disk.  Both episodes are described 
by exponential laws. Their model C does not assume any threshold in the gas density
during the halo/thick-disk phase, and thus allows the formation of the
outer plateau from infalling gas enriched in the halo. 

Figs.~\ref{iron1} and \ref{iron2} show that our model is also able to
reproduce the flat gradient at large Galactocentric distances.  This
is explained in the framework of our chemical evolution model by the
radial dependence of the infall rate (exponentially decreasing with
radius) which, combined with the gas distribution in the halo, results
in a gas falling into the disk rapidly in the regions with
R$\lesssim$12~kpc and at a almost constant rate in the outer
regions (see Fig.~\ref{fig_model1}). In addition, the radial
dependence of the star and cloud formation processes, that are more
efficient in the inner regions, produces a higher SFR in the inner
regions, resulting in a higher metal production. In contrast,
in the outer disk, the low metallicity of the infalling gas and the
low SFR contribute to maintain a flat and slowly evolving metallicity
gradient.

To reproduce a completely flat gradient in the external regions, as
indicated by the older OCs, our model would need an additional
accretion of material uniformly falling onto the disk. This however
would result in an inconsistent behaviour of the current SFR,
predicting higher SFR in the external regions.  We then considered
several possible events that could explain the trend of the
metallicity gradient traced by older OCs. The plateau is most evident
only in the older group of OCs, and thus it could be related to
peculiar episodes in the past history of the Galaxy.  The outer
plateau could be a signature of a past merger at high radii with
pre-enriched material, or similarly to what is suggested by the model of
Chiappini et al.~(\cite{chiappini01}), of a gas halo already enriched.
In this case it would not be necessary to modify the radial profile of
the SFR to have higher element abundance at high radii, since the
inflow of material would bring a higher content in
metals. Alternatively, it could be caused by a strong episodic merger
in the past with clouds/galaxies at high Galactocentric radii.
Sporadic episodes of mergers with galaxies at large Galactocentric
radii, more frequent in the past history of our Galaxy, would not
strongly affect the current SFR and would explain the outer
plateau in metallicity. However, fine tuning of the composition of the
merger would be needed to reproduce the exact plateau value in iron
and the distribution of other elements (see next section).

\subsection{Iron-peak- and $\alpha$-elements}
\label{sect_other} 

\begin{figure}
\psfig{figure=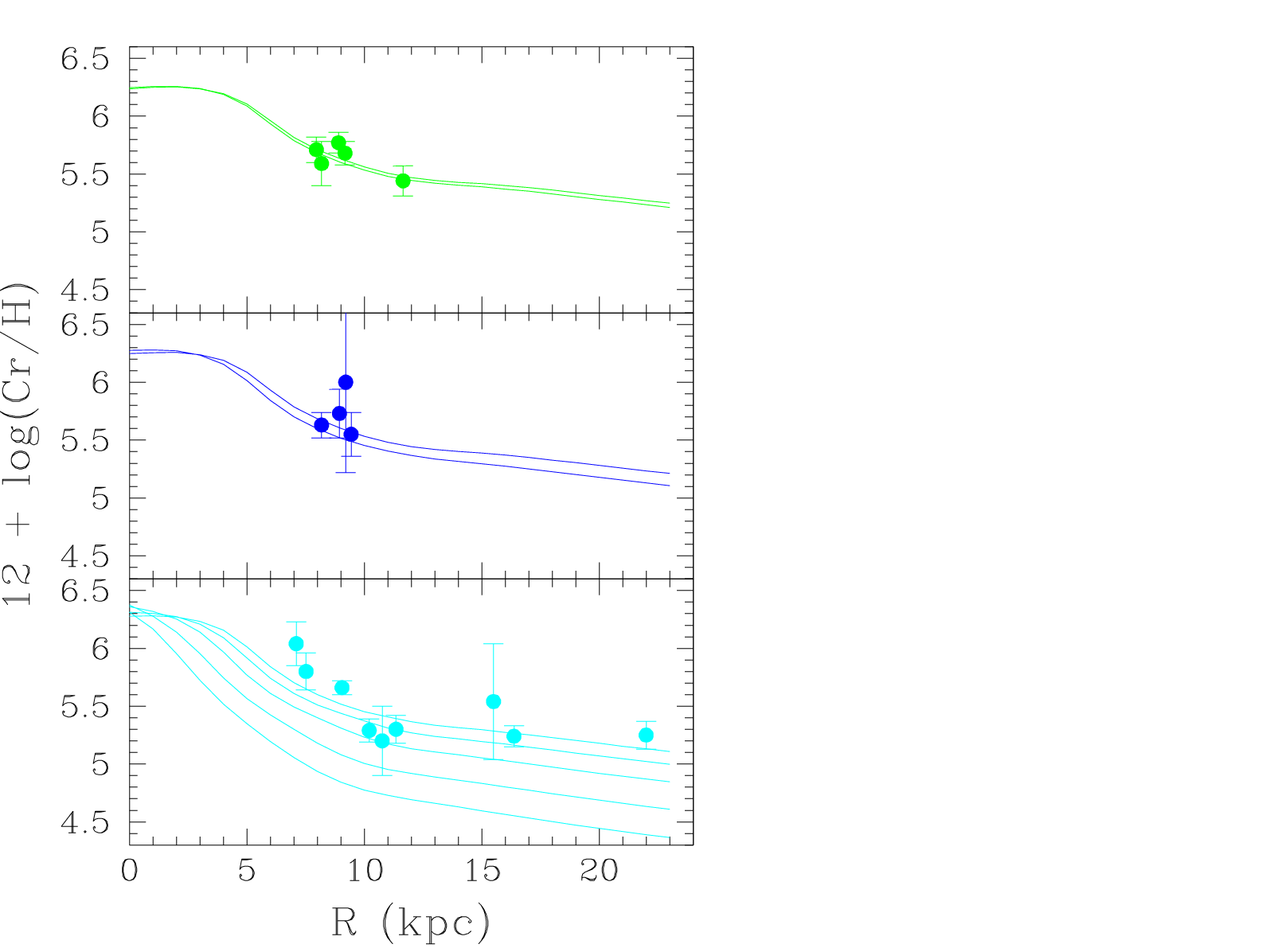, width=11cm, angle=-90}
\psfig{figure=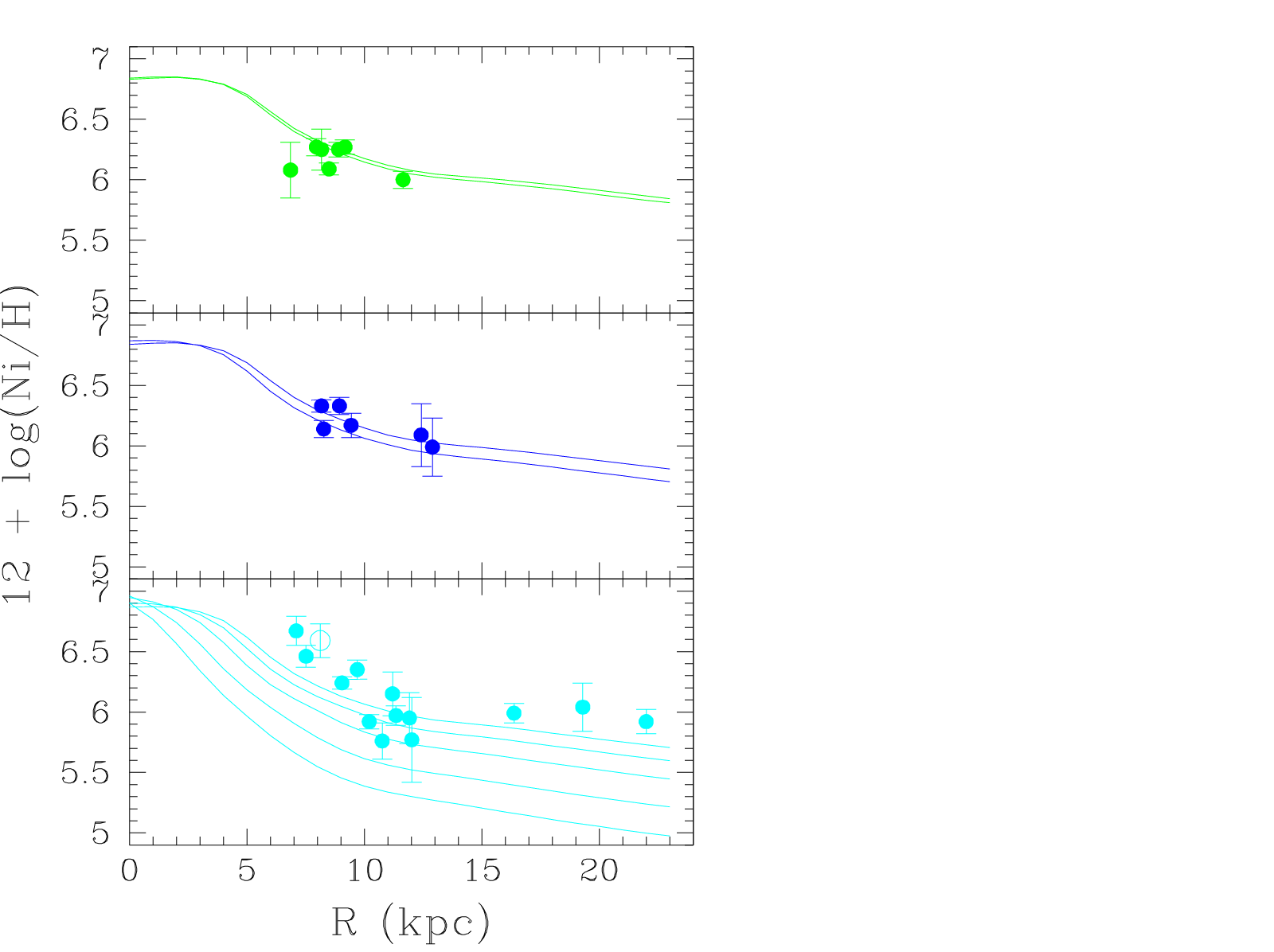, width=11cm, angle=-90}
\caption{Radial gradients of $12+\log({\rm Cr/H})$ and $12+\log({\rm
Ni/Fe})$: comparison between high-resolution data of OCs with our model.  Symbols and line-types are as in
Fig.~\ref{iron1}. These figures are available in electronic form only.}
\label{fepeak}
\end{figure}

\begin{figure} 
\psfig{figure=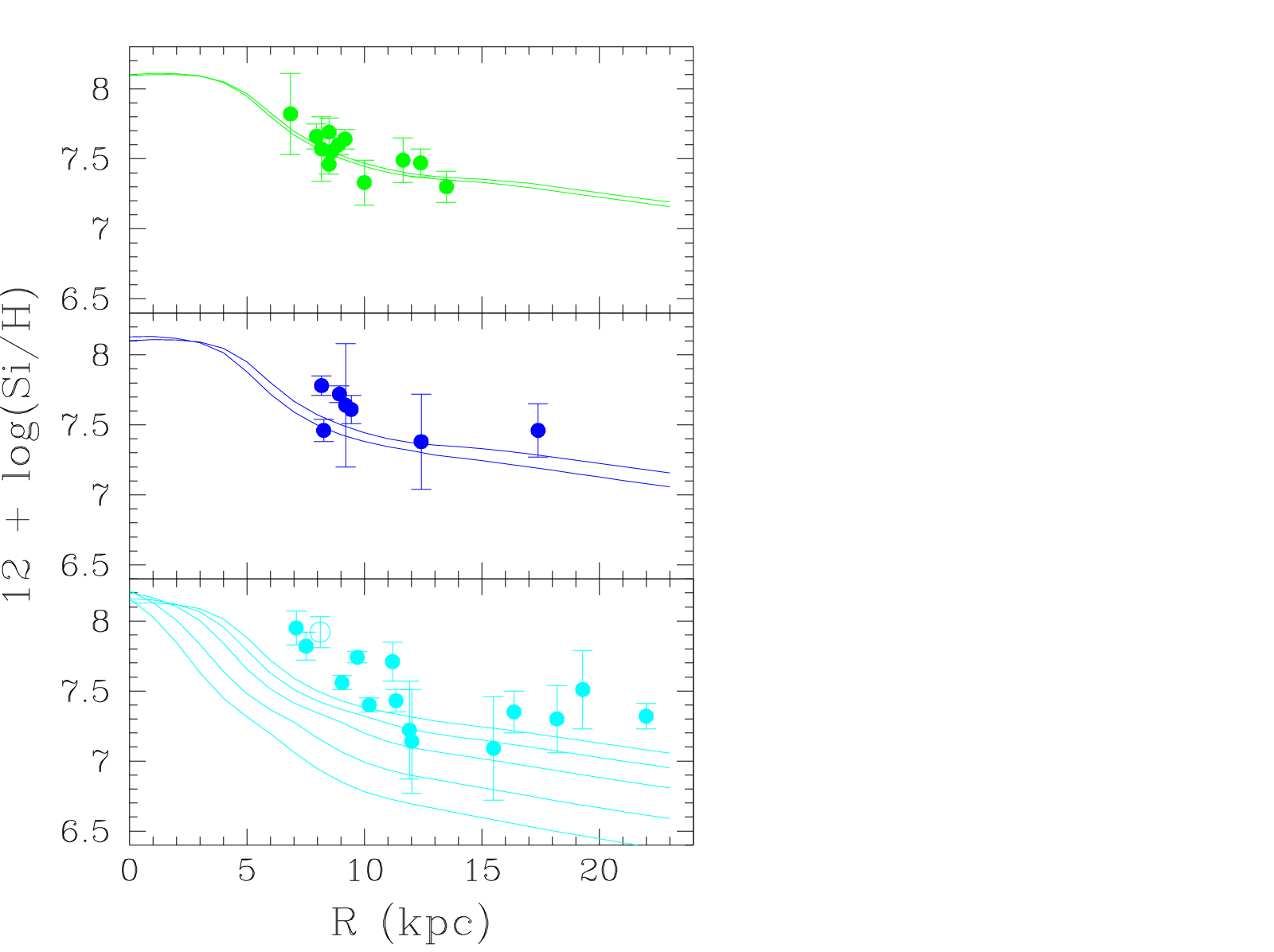, width=11cm, angle=-90}
\psfig{figure=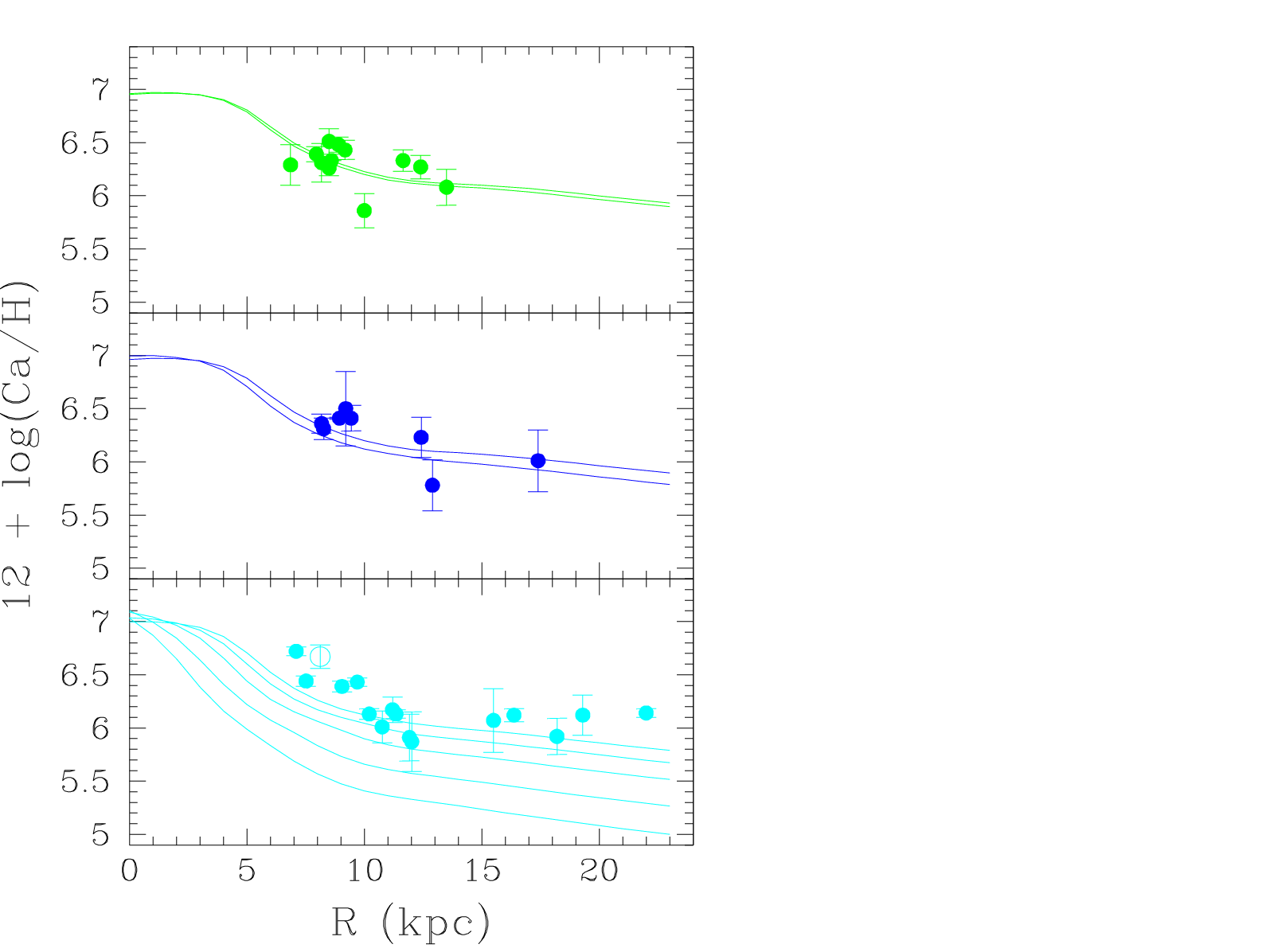, width=11cm, angle=-90} 
\caption{Radial gradients of $12+\log({\rm Si/H})$ and $12+\log({\rm
Ca/H})$: comparison between high-resolution data of OCs with our  model.  Symbols and line-types are as in
Fig.~\ref{iron1}. These figures are available in  electronic form only.}
\label{alpha1}
\end{figure}

\begin{figure} 
\psfig{figure=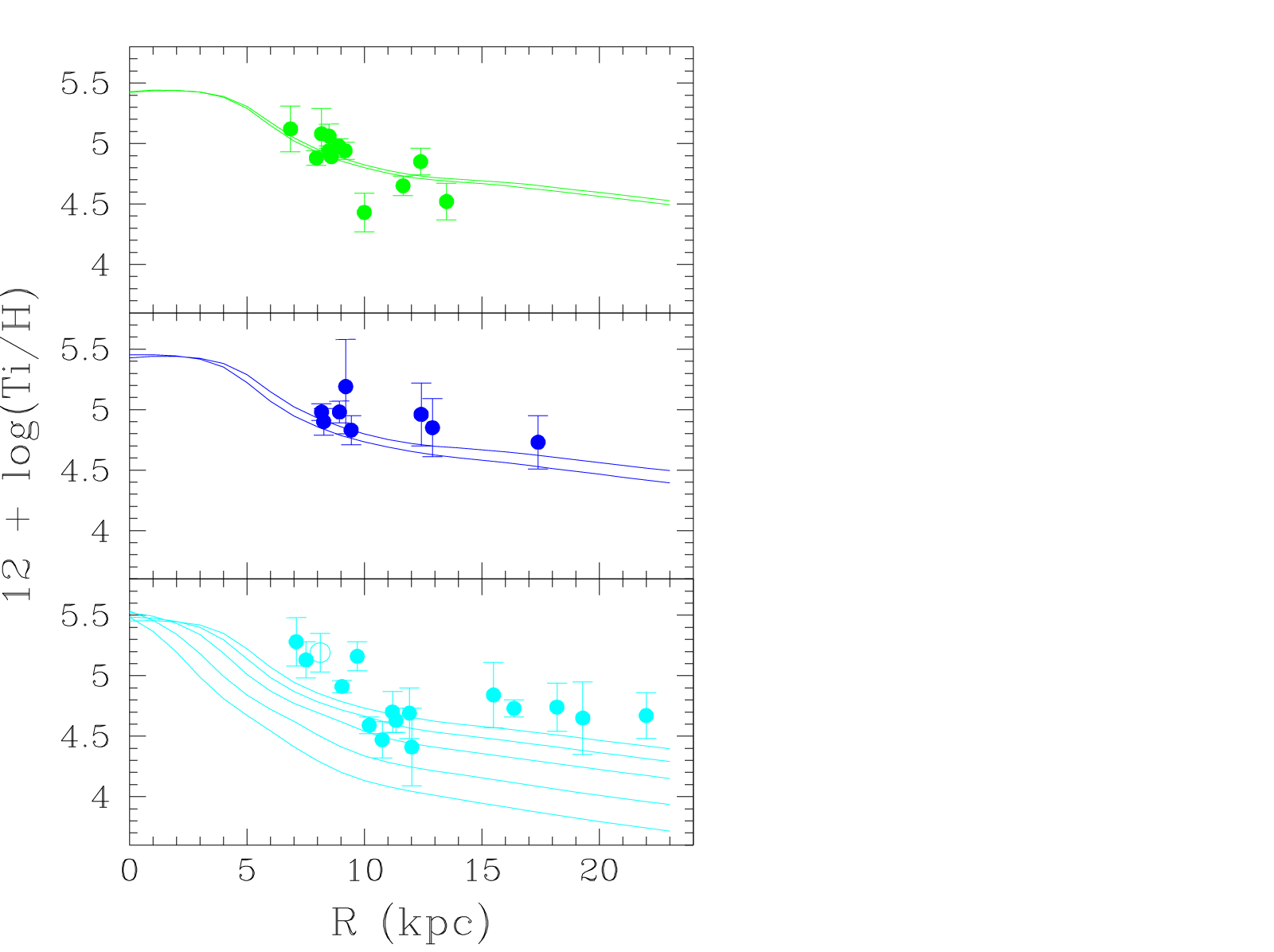,width=11cm, angle=-90}
\caption{Radial gradient of $12+\log({\rm Ti/H})$: comparison
between high-resolution data of OCs with our model decribed.  Symbols and line-types are as in
Fig.~\ref{iron1}. These figures are available in  electronic form only.}
\label{alpha2}
\end{figure}

Two iron-peak elements, Ni and Cr, and three $\alpha$-elements (Si,
Ca, and Ti) have been measured in a large number of OCs belonging to
the selected sample. Their gradients are plotted in Figs.~\ref{fepeak},
\ref{alpha1} and \ref{alpha2}. The radial
gradients are in most cases very similar and also very close to the iron
gradient, with a negative slope at low radii and a slightly decreasing
distribution at larger distances (see Table~\ref{tab_grad_iron}). Note
that in a few cases we were not able to derive a quantitative value
for the slope, due to low statistics and/or to the small range of
R$_{\rm GC}$  covered by the clusters with available abundances.
Also, in two of cases (Ca and Cr) the slope in the 7--12~kpc
range in the intermediate-age bin is consistent with 0, at variance
with that of Fe. For all elements, however, the slope in the oldest age
bin in the 7--12~kpc interval is very similar to the slope of [Fe/H].
This similarity is expected in our model because the metallicity gradient
is determined for all elements by the radial dependence of the SFR and
the infall.

In order to reproduce the absolute mean values of $12+\log({\rm Cr/H})$,
we have assumed a production of Cr from SN~Ia lower by a factor
of $\sim 2$ than that by Iwamoto et al.~(\cite{iwamoto99}), while
[Ni/H], [Ca/H], [Si/H], and [Ti/H] values are all well reproduced
using the original yields by Iwamoto et al.~(\cite{iwamoto99}).
The best agreement of the model prediction with the observed time and
radial behaviours of the Ti abundance is obtained adopting stellar
yields from massive stars higher by a factor of $\sim 2$ than those
computed by Chieffi \& Limongi~(\cite{chieffi04}). This correction is
in agreement with the empirical stellar yields adopted by Fran\c cois
et al.~(\cite{francois04}).

In summary, all elements under consideration, in particular in the most
represented age range, $4~\mbox{Gyr} \lesssim \mbox{age} \lesssim
10$~Gyr, confirm the existence of a steep gradient in the inner
regions and a change of slope at around $\sim$12~kpc with a flattening of
the gradient towards the outer regions.  


\section{[X/Fe] radial gradients}

Although abundance ratios are less model-dependent than absolute
abundances (see e.g. Matteucci \& Chiappini~\cite{matteucci04}), nevertheless they
are affected by assumptions on stellar nucleosynthesis processes, the
IMF, and stellar lifetimes.  In particular, abundance ratios involving
elements produced on short timescales by massive stars dying as SN~II,
like $\alpha$-elements, and elements produced on longer timescales by
e.g.  SN~I, like iron-peak elements, allow us to study the role of
different types of supernovae in the chemical history of the
Galaxy.

\begin{figure}
\psfig{figure=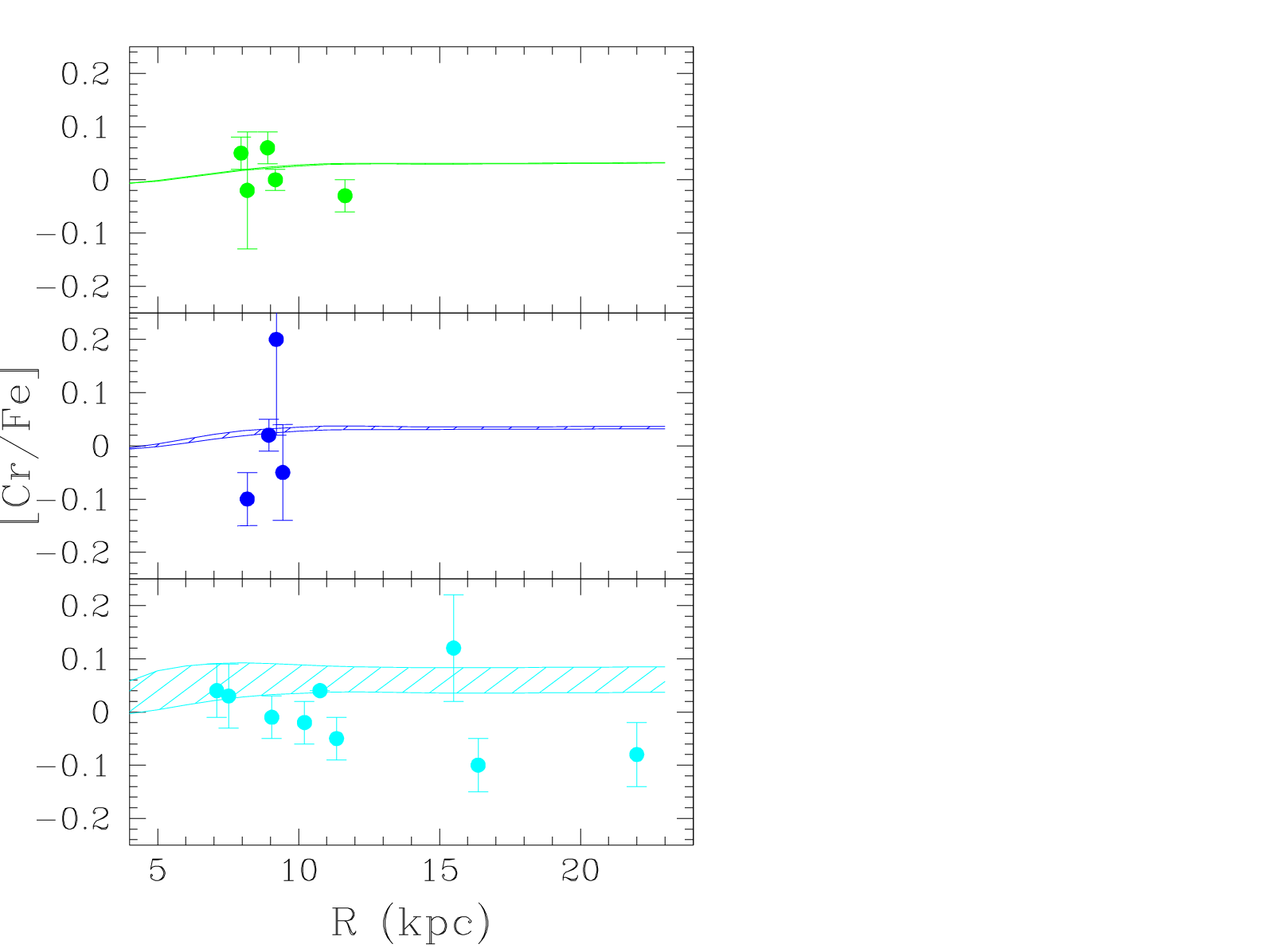, width=11cm, angle=-90}
\psfig{figure=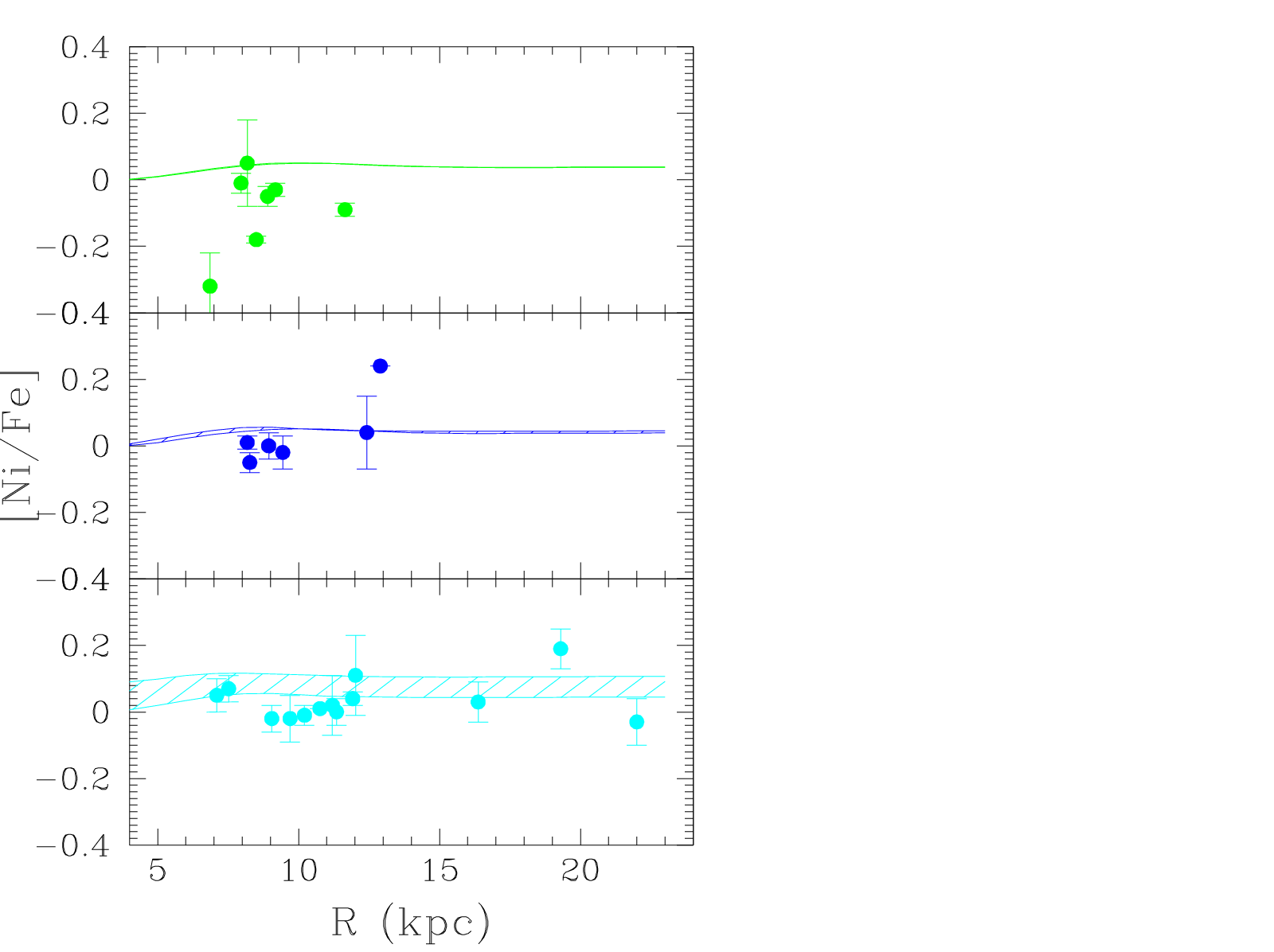, width=11cm, angle=-90}
\caption{Radial gradients of  [Ni/Fe] and [Cr/Fe]: comparison
between high-resolution data of OCs with our model.  
Different panels show abundances for different age bins: top (panel 1), ages $\leq$0.8~Gyr;
middle (panel 2), ages $0.8~\mbox{Gyr} <\mbox{age}\leq 4~\mbox{Gyr}$;
bottom (panel 3), ages $4~\mbox{Gyr} <\mbox{age}\leq 11~\mbox{Gyr}$.  The
models adopted to compare with observations are for 0 and 1 Gyr ago (panel
1), 1 and 4 Gyr ago (panel 2), and 4, 11 Gyr ago (panel 3). The shaded regions indicate the area between the two models 
shown in each panel. These figures are available in electronic form only.} 
\label{fepeakfe}
\end{figure}

\begin{figure} \psfig{figure=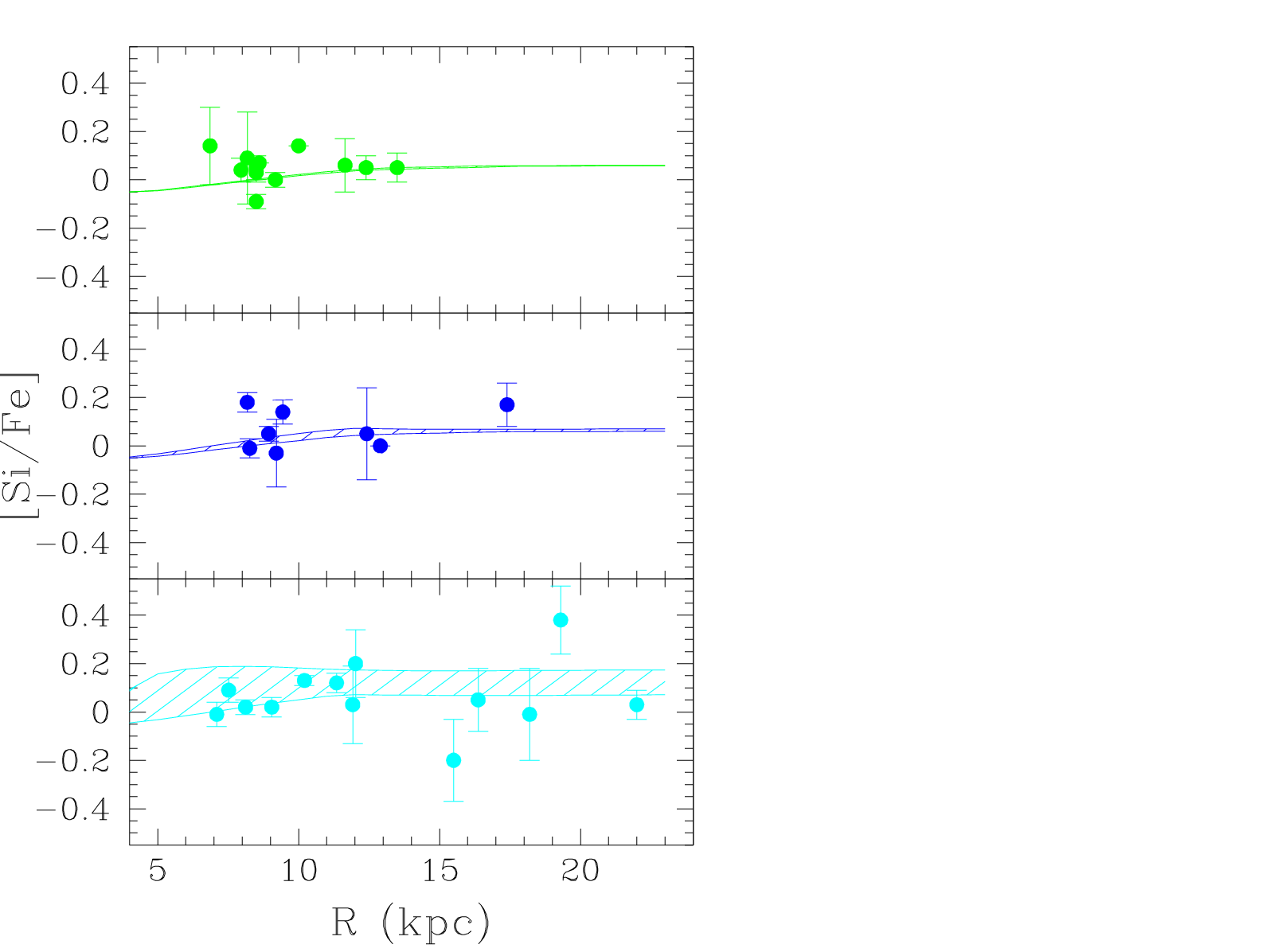, width=11cm, angle=-90}
\psfig{figure=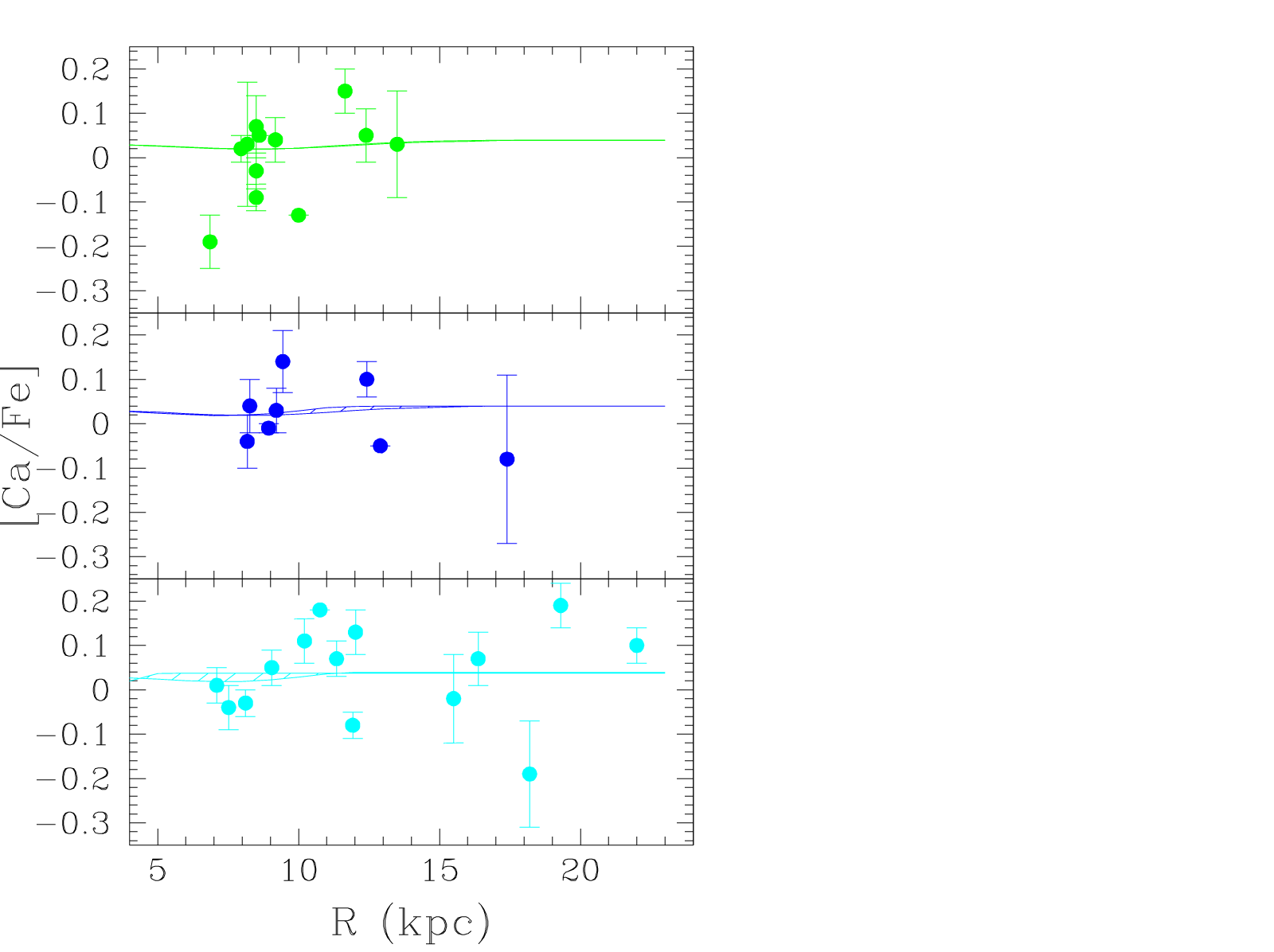, width=11cm, angle=-90} 
\caption{Radial gradients of [Si/Fe] and [Ca/Fe]:
comparison between high-resolution
data of OCs with our model. 
Symbols and line-types are as in Fig.~\ref{fepeakfe}. These figures will be published in the electronic form only.}
\label{alpha1fe}
\end{figure}

\begin{figure} 
\psfig{figure=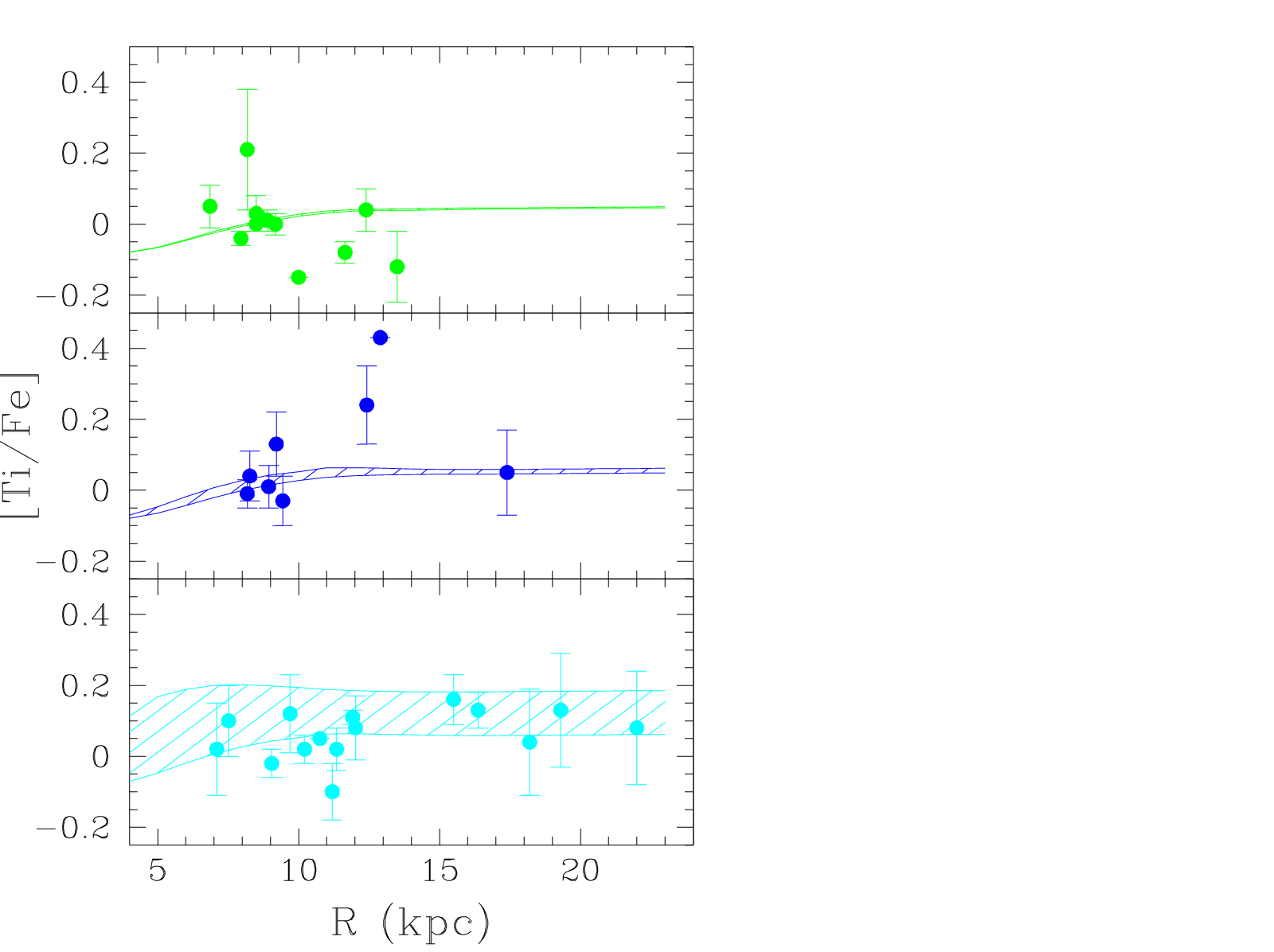,width=11cm, angle=-90}
\caption{Radial gradient of  [Ti/Fe]: 
comparison between high-resolution
data of OCs with our model. 
Symbols and line-types are as in Fig.~\ref{fepeakfe}. These figures are available in electronic form only.}
\label{alpha2fe}
\end{figure}

In Fig.~\ref{fepeakfe}, \ref{alpha1fe}, and
\ref{alpha2fe}, we compare the values [X/Fe] from high-resolution data  with the predictions of
the model. 
The production of iron-peak elements is dominated by SN~Ia. Since iron
has the same SN~Ia origin, the ratios [X/Fe] of these elements are not
expected to show any evident temporal evolution.  

Two iron-peak elements, Ni and Cr, have been measured in a 
large number of clusters belonging to our sample.
Both elements show an almost constant average value, close to the
solar one at all ages.  The absolute mean values of [Cr/Fe] are
reproduced assuming a production of Cr from SN~Ia lower by a factor
of $\sim 2$ than that by Iwamoto et al.~(\cite{iwamoto99}), while
[Ni/Fe] values are well reproduced using the original yields by
Iwamoto et al.~(\cite{iwamoto99}).

Models and observations of both elements, in particular in the most
represented age range, $4~\mbox{Gyr} \lesssim \mbox{age} \lesssim
11$~Gyr, confirm a constant behaviour of [X/Fe] with Galactocentric
radius.  
We recall that typical formal uncertainties in [X/Fe] vary from 0.01 to 0.15 dex.
In the case of Ni lower values of [Ni/Fe] are observed in
the few young clusters analyzed than in the older cluster sample. 
Also the  [Cr/Fe] of some clusters in the old-age bin have lower values than expected from the model. 
Both these behaviours are not reproduced by the model, and they could be statistical variations, not significant, 
due to the small sample of clusters with these measurements.

Data for the three $\alpha$-elements observed in the high-resolution
sample of OCs (Si, Ca, and Ti) indicate a rather flat
distributions with Galactocentric distance, and nearly solar
[$\alpha$/Fe] values, slightly higher for older clusters.  The
largest contribution to the production of Si, Ca, and Ti comes from
massive stars with $M>8 M_\odot$ (e.g., Chieffi \& Limongi
\cite{chieffi04}) while small amounts of these elements are also
produced by SN~Ia (e.g., Iwamoto et al.~\cite{iwamoto99}).  This is
especially true for Ca: for example, SN~Ia models by Iwamoto et
al.~(\cite{iwamoto99}) predict a typical synthesized mass of Ca of
about 0.02--0.03~$M_\odot$ per SN~Ia, to be compared with the
$0.016~M_\odot$ of Ca produced by a star of 25~M$_\odot$ of solar
metallicity (cf. Chieffi \& Limongi \cite{chieffi04}). Thus the
production of Ca and Fe, integrated over the IMF, occur on comparable
time scales, and the [Ca/Fe] gradient does not show any evident
temporal evolution. Silicon, on the other hand, being mostly produced by 
rapidly evolving massive stars, shows high values compared to Fe in
the past, and a more rapidly increasing [Si/Fe] radial
gradient than e.g., the [Ca/Fe] gradient, as suggested by e.g., Cescutti
et al.~(\cite{cescutti06}).  These temporal and spatial behaviours are
well reproduced by the model, that also predicts a slight enhancement
of [$\alpha$/Fe] in the outer regions, due to the different spatial
location of SN~II and SN~Ia.  This is partially confirmed by the
limited number of high-resolution observations of OCs.

Titanium shows a large dispersion with cluster age, similarly to the
one observed for [Si/Fe].  The best agreement of the model prediction
with the observed temporal and radial behaviours of the Ti abundance
is obtained adopting stellar yields from massive stars higher by a
factor of $\sim 2$ than those computed by Chieffi \&
Limongi~(\cite{chieffi04}).  This correction is in agreement with the
empirical stellar yields adopted by Fran\c cois et
al.~(\cite{francois04}).



As described in the Appendix, with our adopted
model parameters (infall, star and cloud formation efficiencies,
chemical yields, etc.)  the best agreement of the theoretical
gradients with the observations is obtained using the IMF by Kroupa et
al.~(\cite{kroupa93}).  We also checked several other parameterizations of
the IMF: assuming the ``best'' values (i.e. measured 
independently) for the various efficiencies of star and cloud formation at  solar radii, 
we have varied the adopted parametrization of the IMF.   
A good agreement was also obtained with the IMFs by Ferrini et
al.~(\cite{ferrini90}) and Scalo~(\cite{scalo86}).  These parameterizations
are characterized by  a lower number of massive stars when compared e.g., with
the Salpeter~(\cite{salpeter55}), Scalo~(\cite{scalo98}), and Chabrier
et al.~(\cite{chabrier03}) IMFs.  The high-resolution OC abundances
of $\alpha$-elements seem to weaken the use of  these other parameterizations
that in fact result in an overproduction of massive stars, and,
consequently, a  production of $\alpha$-elements greater than observed.
For example, a Salpeter IMF would result in an overproduction of 
Ti in the model of $\sim 0.3$~dex with respect to the 
observations.

\section{The time evolution of the metallicity gradient}
\label{sect_discussion}
 
\setcounter{table}{1}
\begin{table} 
\caption{Average [Fe/H] from OC data of  different epochs and radial
regions (see the text for the definition of radial regions).}
\label{tab_average}
\centering
\begin{tabular}{rrr}
age (Gyr)  & Region II            & Region III \\
\hline
$<0.8$       & $0.016 \pm 0.14$         & $-0.17 \pm 0.12$   \\
0.8--4       & $0.05 \pm 0.11$          & $-0.29 \pm 0.16$   \\
4--10         & $-0.04 \pm 0.29$         & $-0.33 \pm 0.13$   \\
\hline 
\end{tabular}
\end{table}
%

In their paper, Maciel et al. (2006) showed the results on the
time evolution of the [Fe/H] gradient obtained by several authors,
both from models (Hou et al.~\cite{hou00}, Chiappini et
al.~\cite{chiappini01}) and from observations (PNe by Maciel et
al.~2006, OCs by Friel et al.~2002, Chen et al.~2003, and \hii\
regions, Cepheids, OB stars from several authors, see references
therein).  The metallicity gradient was computed over the whole disk. 
However, as already
remarked, the gradient does not have a constant slope over
the whole disk, and, thus, a direct comparison with our data is difficult to
obtain.

Therefore we compared our results with the samples of OCs by Friel et
al.~(2002) and Chen et al.~(2003) in the same radial regions.  Their
samples were discussed in the previous sections. In summary,
they differ from our sample in data quality (high-resolution
spectroscopy vs. low-resolution spectroscopy and/or photometry) and
 radial range (they extend up to $\sim$16~kpc, while our sample extends
up to 22~kpc).  The OCs located at high Galactocentric radius are
fundamental in determining the slope in the older and intermediate age
bin.  In order to compare with the results of Friel et al.~(2002) and
Chen et al.~(2003) we have re-computed the slope of the older and
intermediate age bins for $R_{GC}\leq16$, obtaining a slope for the
cumulative sample of -0.067~dex kpc$^{-1}$, and for  only the old bin
of -0.079~dex kpc$^{-1}$, in closer agreement with their results:
-0.075~dex kpc$^{-1}$ for the sample of Chen et al.~(2003) with ages
$>$0.8~Gyr and -0.075~dex kpc$^{-1}$ for the sample of Friel et
al.~(2002) with ages $>$4~Gyr.  This points out the importance of OCs
at larger Galactocentric distances for a correct determination of the
outer gradient.

The comparison with PNe is more tricky.  The accuracy of the Galactic
metallicity gradient derived for PNe has  been debated for a
long time because of the large uncertainties on PN distances.  Recent
results (Stanghellini et al.~\cite{stanghellini06}, Perinotto \&
Morbidelli~\cite{perinotto06}) have shown very different gradients for
the Galactic PN population than previous results.  At variance with
Maciel et al.'s results, Stanghellini et al.~(\cite{stanghellini06}) and  Perinotto \&
Morbidelli~(\cite{perinotto06}) found an almost flat gradient at all
epochs.  
On the other hand, some other recent works find 
steeper gradients (e.g., Pottasch \& Bernard-Salas~\cite{pottasch06};
Maciel \& Costa~\cite{maciel08}) pointing to the need for an accurate determination of
the age in order to measure gradients. The main differences among
these works are the adopted distance scales, demonstrating how this
choice can affect the estimate of the metallicity gradient,  in
conjunction with different age estimates for the PN progenitors.  Thus
a faithful determination of the Galactic metallicity gradient from PNe
is far from  being obtained.

A more appropriate way to compare the chemical evolution model
with observations is to divide the Galaxy in several radial
regions.  In Tables~\ref{tab_average} and \ref{table:grad} we list
the the average abundances of iron in each age bin and radial region,
and the best-fit slopes in the three radial regions. The latter were
computed through a weighted fit procedure for the OCs and taken from
Andrievsky et al. (\cite{andri04}) and Lemasle et
al. (\cite{lemasle07}) for the Cepheids.  The slopes of the models are
also shown.  As anticipated above, three age bins were considered: age
less than or equal to 0.8~Gyr, in the first panel; $0.8<\mbox{age}\leq
4$~Gyr, in the second panel; and $4<\mbox{age}\leq 11$~Gyr in the
third panel.  For OCs with ages between 4~Gyr and 11~Gyr, we excluded
from the calculation of gradient the metal-rich cluster NGC~6791
(marked with an empty circle in Figs.~\ref{iron1}, \ref{fepeak},
\ref{alpha1},\ref{alpha2}), because it was probably born in a
different place in the disk or could even have been captured by our
Galaxy during a merger event (Carraro et al.~\cite{carraro06}, but see
also Bedin et al.~\cite{bedin06} for a contrasting view). Thus the age
range for our further analysis is between 4~Gyr and 10~Gyr.

\begin{figure} 
\psfig{figure=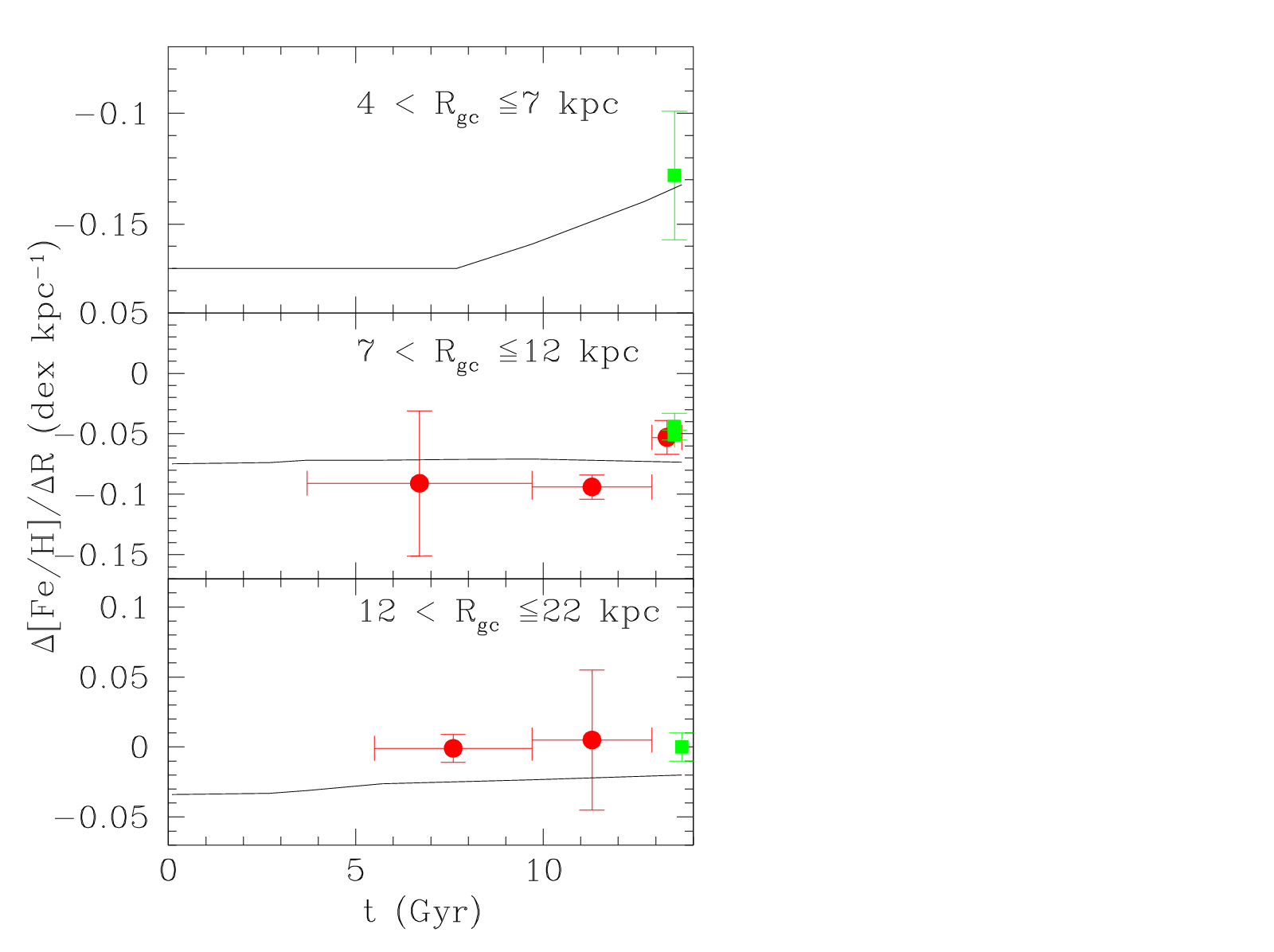,width=11cm, angle=-90}
\caption{Time evolution of the slope of the iron gradient in the three
radial region analyzed: $4$~kpc $\lesssim R_{\rm{GC}}\lesssim7$~kpc
(region~I), $7$~kpc $\lesssim R_{\rm{GC}}\lesssim12$~kpc (region~II),
$12$~kpc $\lesssim R_{\rm{GC}}\lesssim22$~kpc (region~III).  Continuous
lines are the slopes of the model gradients approximated with a linear
fit.  Filled squares are the slopes of the gradients of  Cepheids.
Filled circles represent the slopes of OC gradients with their uncertainties in
the three age bins, as in Table~\ref{tab_grad_iron}.}
\label{slopes}
\end{figure}

First, we note that, although it is well known that a strict
age-metallicity relationship does not hold in the Galactic disk, a
tendency is present for older OC to have on average lower metallicities.
Second,
each radial
region shows a different behaviour due to the distinct ratios of SFR
and infall. To better analyze the time evolution of the metallicity
gradients, we plot in Fig.~\ref{slopes} the slopes of the gradient
for the specific case of iron, the best studied element, in the three
radial regions defined above. Similar behaviours are observed also for
the other elements analyzed in this study.

As there are no observations of OCs in region~I, it is not possible to
obtain the time evolution of the gradient in this region.  In
region~II, our sample of OCs does not show any conclusive evidence for
flattening with time of the slope of the gradient, although the
present time gradient seems slightly flatter then the gradient in the
two older age bins.  Finally, in region~III, the OC metallicities
produce a gradient consistent with a zero slope at all epochs.

Our data have allowed us to consider the time evolution in different
regions of the disk, showing that in the $\sim 7$--22~kpc interval no
marked evolution is present.

The observations of Cepheids by Andrievsky et al.~(\cite{andri04})
(see Table~\ref{tab_grad_iron} and filled squares in
Fig.~\ref{slopes}) show a good agreement with the absolute [Fe/H]
predicted by the model in regions~I, II, and III at the present time
(see Table~\ref{tab_grad_iron}).  In region~I, the model predicts a
flattening of the metallicity gradient from the past epochs to the
present time that cannot be tested because only Cepheid data are
available in this region; in regions~II and III it predicts an overall
increase of metallicity at all radii but no significant change in the
slope of the gradient (see Fig.~\ref{slopes}).


This can be understood as follows. In our model, the inner part of the
disk is built by the rapid collapse of the halo, whereas the outer
parts by the continuous slow accretion of intergalactic gas.  In the
inner disk, a steep chemical gradient is initially established by the
strong concentration of infalling gas (see eq.~\ref{eq_infall}), on a
timescale probably of the order of a few Gyr or less (OCs are not able
to constrain the chemical evolution of the gradient for the first
$\sim 5$~Gyr of evolution of the Galaxy). 
The  model predictions indicate that this
initial gradient is preserved in the subsequent evolution of the disk.

Turbulent mixing of the interstellar gas seems to be inefficient in
smoothing the initial metallicity gradient over the lifetime of the
Galaxy, in agreement with theoretical estimates; for example,
according to Talbot \& Arnett~(\cite{talbot73}), the timescale for
mixing a region of a radial extent of 5 kpc like Region~II is about
$75$~Gyr. Also the role of large-scale radial flows, often invoked in
the chemical evolution of disk galaxies, seems quite marginal, since
their effect is to steepen the metallicity gradient (Lacey \&
Fall~\cite{lacey}; Tosi~\cite{tosi88}; Portinari \&
Chiosi~\cite{portinari00}), contrary to observational evidence.
However Friedli (\cite{friedli98}) found that flows induced by bars
can flatten the gradient.  In the outer disk, on the other hand,
accretion of intergalactic gas and disk evolution occur almost
independently of radius, and the metallicity gradient remains flat.

We note that the slope of older OCs is significantly affected by the
presence of a number of old clusters located at low Galactocentric
distances with solar or super-solar metallicities. Old clusters have
had enough time to move along their orbits, and the possibility exists
that they were actually born in inner regions far from their present
location. Although it seems unlikely that all old clusters with
(over)-solar metallicities have different birthplaces, we stress that
knowledge of the orbits of these older OCs, and thus their previous
location, would help in definitively settling the issue. Also, a few
clusters located at distances of 11-12 kpc are present with [Fe/H]
somewhat below the plateau value observed at larger Galactocentric
distances and characterized by large error bars. New observations for
these clusters should be obtained in order to better constrain the
slope of the gradient.  Finally, metallicity determinations should be
performed for both young and old clusters located at Galactocentric
distances lower than $\sim 7$~kpc.

\section{Summary} 
\label{sect_summary}
s
We have assembled a homogeneous sample of abundances in OCs from
high-resolution spectroscopic data and compared it  to the results of
a GCE model.  We examined in particular the radial gradients of Fe/H,
Si/H, Ca/H, Ti/H, Ni/H, and Cr/H, and also their [X/Fe]
gradients at different epochs during the lifetime of the Galaxy. We
have divided our data into three age bins and three radial
regions. Our main results are the following:


\begin{itemize}

\item[({\em i}\/)] From the observations of OCs and Cepheids (see
Table~\ref{tab_grad_iron}), we find that the metallicity gradients of
all the elements analyzed are very steep at low Galactocentric radii,
in particular for $R_{\rm GC}\lesssim 7$~kpc.  The gradients show a
change of slope around 7~kpc, and further flattening at about 11--12~kpc,
resulting in a metallicity plateau in the outer regions.

\item[({\em ii}\/)] The combination of the observations of Cepheids and
OCs allow us to follow the time evolution of the metallicity gradient
and to compare it with the results of a GCE model. The time evolution
of the gradient can be followed where observations of OCs with
different ages are available for $R_{\rm GC}> 7$~kpc.  At
Galactocentric radii $7$~kpc $\lesssim R_{\rm GC}\lesssim 12$~kpc the
combined observations of Cepheids and OCs suggest that the slope
of the metallicity gradient has not changed significantly over the past
$\sim 10$~Gyr; if anything, the gradient has flattened with time, as
also suggested by the model for $R_{\rm GC} < 7$~kpc. For
$12~\mbox{kpc} \lesssim R_{\rm GC} \lesssim 22$~kpc, OCs belonging to
different epochs and Cepheids all suggest a flat metallicity
distribution.

\item[({\em iii}\/)] 
The GCE model that better reproduces the
data is characterized by an inside-out formation of the disk. The
infall of gas is represented by an exponential law which, combined
with the distribution of gas in the halo, produces a rapid collapse in
the regions with R$\lesssim$12~kpc and a uniform accretion in the
outer regions.  The inner regions are rapidly evolving due to the
higher infall and SFR, while the outer parts evolve more slowly.  The
metallicity in the outer disk increases uniformly at all radii without
a marked change of slope of the gradient.  Peculiar episodes of
mergers in the past history of the Galaxy would explain the outer
metallicity plateau indicated by the older clusters.
\end{itemize}

\begin{landscape}
\setcounter{table}{0}
\begin{table*} 
\scriptsize
\caption{Open clusters included in our sample. We list in Col.~1
the cluster name, in Col.~2 the cluster age, in Cols.~3 and 4 the
Galactocentric radius and its reference (an asterisk indicate that we
computed R$_{\rm GC}$  assuming the cluster distance given in the quoted
reference). In Cols.~5, 6, 7, 8, 9, and 10 we list the main properties
of the observations and of the analysis: the signal to noise ratio, the
resolving power, the number and type of stars analyzed, the adopted method
(Equivalent widths, EW, or spectral synthesis), the iron abundance with
its error, and the reference.}
\begin{tabular}{llllllllll}
\hline
\hline
Cluster &Age            & $R_{\rm{GC}}$ &  Ref.              & S/N      & R     &N.Stars&Method   &[Fe/H]  & Ref.\\
         & (Gyr)         &(kpc)         & ($R_{\rm{GC}}$)&          &       &      &         &        & ([Fe/H])     \\
\hline  
IC4665   & 0.025        &8.18         & Manzi et al.~(\cite{manzi08})$^*$        &   30-150    & 60,000       &18,dwa     &Synt       & $-0.03\pm0.04$ & Shen et al..~(\cite{shen})\\ 
IC2602   & 0.035        &8.49         & WEBDA$^*$\footnote{http://www.univie.ac.at/webda/} & 30-100 & 18,000-44,000&9,dwa     &EW       & $-0.05\pm0.05$ &Randich et al. (2001)              \\
IC2391   & 0.053        &8.50         & WEBDA$^*$                                &   30-100 & 18,000-44,000&4,dwarfs  &EW       & $-0.03\pm0.07$ &Randich et al. (2001)                \\ 
Blanco 1 & 0.1-0.15     &8.50         &   Friel~(\cite{friel06})                 &   70-100 & 50,000       &8,dwa     &EW       & $+0.04\pm0.04$ & Ford et al.~(\cite{ford05})   \\           
Pleiades & 0.125        &8.60         &   Friel~(\cite{friel06})                 &   70     & 45,000       &15,dwa    &EW       & $-0.03\pm0.02$ & Boesgaard \& Friel~(\cite{boesgaard90})       \\ 
NGC6475  & 0.220        &8.22         & Robichon et al.~(\cite{robichon99})$^*$  &   50-150 & 29,000       &13,dwa    &EW       & $+0.14\pm0.06$ &Sestito et al. (\cite{sestito03})                 \\
M11      &  0.250       &6.86         & Gonzalez \& Wallerstein~(\cite{gonzalez00})$^*$&  85,130 & 38,000 &10,gia     &EW/synthesis& $+0.17\pm0.13$ &Gonzalez \& Wallerstein (2000)\\      
M34      & 0.250        &8.90         &   Friel~(\cite{friel06})                 &   70     & 45,000       &4,dwa     &EW       & $+0.07\pm0.03$ & Schuler et al.~(\cite{schuler03})       \\
NGC3960  & 0.7          &7.96         & Friel et al.~(\cite{friel02})            &  100-150 &45,000        &6,giants  &EW       &$+0.02\pm0.04$ & Sestito et al.~(\cite{sestito06})\\ 
Hyades   & 0.7          &8.50         &   Friel~(\cite{friel06})                 &   100-200& 60,000       &55,dwa    &EW       & $+0.13\pm0.05$ & Paulson et al.~(\cite{paulson03})       \\ 
Praesepe & 0.7          &8.62         &   Friel~(\cite{friel95})                 &   130     & 100000       	   &7	      &EW	& $+0.27\pm0.05$ & Pace et al. (\cite{pace08})   \\    
NGC2324  &0.7           & 11.65       & Friel et al.~(\cite{friel02})            &   90-150 &45,000        &7,gia     &EW       &$-0.17\pm0.05$ & Bragaglia et al.~(\cite{bragaglia07})\\ 
NGC2660  & 0.8          &9.18         & Friel~(\cite{friel95})                   &   50-80  &45,000        &5,gia     &EW       &$+0.04\pm0.04$ & Sestito et al.~(\cite{sestito06})\\ 
Mel71    & 0.8          &10.0         &   Friel~(\cite{friel06})	         &   100    & 16,000,34,000&2,gia     &EW       & $-0.32\pm0.16$ & Brown et al.(1996)               \\    
Ru 4     &0.8           & 12.40       &  Carraro et al.~(\cite{carraro07})$^*$   &25--40    & 40,000       &3,gia     &EW       &$-0.09\pm0.05$ &Carraro et al.~(\cite{carraro07})  \\
Ru7      & 0.8          &13.50        &  Carraro et al.~(\cite{carraro07})$^*$   &25--40    & 40,000       &5,gia     &EW       &$-0.26\pm0.05$&Carraro et al.~(\cite{carraro07})  \\
NGC2360  & 0.9          &9.28         &   Friel et al.~(\cite{friel02})          &   50-100 & 28,000       &4,gia     &EW       & $+0.07\pm0.07$ & Hamdani et al.~(\cite{hamdani00}) \\
NGC2112  & 1.0          &9.21         &   Friel et al.~(\cite{friel02})          &   70-80  & 16,000,34,000&2,gia     &EW       & $+0.16\pm0.03$ & Carraro et al.~(\cite{carraro08})           \\
NGC2477  & 1.0          &8.94         & Friel et al.~(\cite{friel02})            &  100-150 &45,000        &6,gia     &EW       &$+0.07\pm0.03$ & Bragaglia et al.~(\cite{bragaglia07})\\ 
NGC7789  &1.1           &9.44         &   Friel et al.~(\cite{friel02})          &   $>$50  & 30,000       &9,gia     &EW/synthesis& $-0.04\pm0.05$ &Tautvaisiene et al.~(\cite{taut05})   \\
NGC3680  & 1.4          &8.27         &   Friel et al.~(\cite{friel02})          &    80    & 100,000      &2,dwa     & EW&     $-0.04\pm 0.03$  &   Pace et al.~(\cite{pace08})\\
NGC6134  & 1.6          &7.68         & WEBDA$^*$                                &   80-200 & 43,000       &3,gia     &EW       & $+0.15\pm0.07$ & Carretta et al.~(\cite{carretta04})\\  
IC4651   & 1.7          &7.70         &    Friel~(\cite{friel06})                &   100    & 48,000       &4,gia     &EW       & $+0.11\pm0.01$ & Carretta et al..~(\cite{carretta04})\\ 
Be73     &1.9           &17.40        &   Carraro et al.~(\cite{carraro07})$^*$  &25--40    & 40,000       &2,gia     &EW       &$-0.22\pm0.10$&Carraro et al.~(\cite{carraro07})  \\
NGC2506  &2.2           &10.88        &   Friel et al.~(\cite{friel02})          &   80-110 & 48,000       &3,gia     &EW       & $-0.20\pm0.02$ & Carretta et al.~(\cite{carretta04})   \\
NGC2141  &2.5           &12.42        &   Friel et al.~(\cite{friel02})          &   100    & 28,000       &8,gia     &EW       & $-0.18\pm0.15$ & Yong et al.~(\cite{yong05})   \\
NGC6819  & 2.9          &8.18         &   Friel et al.~(\cite{friel02})          &   130    & 40,000       &3,gia     &EW       & $+0.09\pm0.03$ &Bragaglia et al. (2001)            \\ 
Be66     &3.7           &12.9         &   Phelps \& Janes~(\cite{phelpsjanes96}$^*$)&   5,15& 34,000       &2,gia     &EW       & $-0.48\pm0.24$ & Villanova et al.~(\cite{villanova05})   \\
Be22     &4.2           &11.92        &   Friel~(\cite{friel95})                 &   20-25  & 34,000       &2,gia     &EW       & $-0.32\pm0.19$ & Villanova et al.~(\cite{villanova05})   \\ 
M67      & 4.3          &9.05         &   Friel et al.~(\cite{friel02})          &   90-180 & 45,000       &10,dwa    &EW       & $+0.03\pm0.01$ & Randich et al.~(\cite{randich06})\\    
Be20     &4.3           & 16.37       & Friel et al..~(\cite{friel02})           &   40-80  &45,000        &2,gia     &EW       &$-0.30\pm0.02$ & Sestito et al.~(\cite{sestito08})\\ 
Be29     &4.3           & 22.0        & Friel~(\cite{friel06})                   &   25-150 &45,000        &6,gia     &EW       &$-0.31\pm0.03$ & Sestito et al.~(\cite{sestito08})\\ 
NGC7142  &4.4           &  9.70       & Jacobson et al.~(\cite{jacobson08})      &   100    &30,000        &4,gia     &EW       & $0.14\pm0.01$ & Jacobson et al.~(\cite{jacobson08})  \\
NGC6253  &  4.5         &7.10         & Bragaglia \& Tosi~(\cite{bragagliatosi06})&   60-140 &45,000       &5,gia     &EW       &$+0.36\pm0.07$ & Sestito et al.~(\cite{sestito07})\\ 
NGC2243  &4.7           &10.76        &   Friel et al.~(\cite{friel02})          &   100    & 30,000       &2,gia     &EW       & $-0.48\pm0.15$ & Gratton \& Contarini~(\cite{gratton94})  \\ 
Be75     & 5.1          &15.50        &  Carraro et al.~(\cite{carraro07})$^*$   &25--40    & 40,000       &1,gia     &EW       &$-0.22\pm0.20$&Carraro et al.~(\cite{carraro07})  \\
Be31     &5.2           &12.02        &   Friel et al.~(\cite{friel02})          &   100    & 28,000       &5,gia     &EW       & $-0.57\pm0.23$ & Yong et al.~(\cite{yong05})     \\    
Mel66    & 5.4          &10.21        &   Friel et al.~(\cite{friel02})          &   80-130 &45,000        &6,gia     &EW       &$-0.33\pm0.03$ & Sestito et al.~(\cite{sestito08})\\ 
Be32     &6.1           &11.35        & Friel et al.~(\cite{friel02})            &  50-100  &45,000        &9,gia     &EW       &$-0.29\pm0.04$ & Sestito et al.~(\cite{sestito06})\\ 
Be25     &6.1           &18.20        &  Carraro et al.~(\cite{carraro07})$^*$   &25--40    & 40,000       &4,gia     &EW       &$-0.20\pm0.05$&Carraro et al.~(\cite{carraro07})  \\
NGC188   & 6.3          &9.35         &   Friel et al.~(\cite{friel02})          &   20-35  & 35,000-57,000&5,dwa     &EW       & $+0.01\pm0.08$ &Randich et al. (2003)                 \\
Saurer1  &6.6           &19.3         &   Friel~(\cite{friel06})                 &   80     & 34,000       &2,gia     &EW       & $-0.38\pm0.14$ & Carraro et al.~(\cite{carraro04})   \\  
Cr 261   & 8.4          &7.52         & Friel et al.~(\cite{friel02})            &   70-130 & 45,000       &7,gia     &EW       &$+0.13\pm0.05$ & Sestito et al.~(\cite{sestito08})\\ 
Be17     &  10.07       &11.2         &   Friel et al.~(\cite{friel05})          &   15-100 & 25,000       &4,gia     &EW       & $-0.10\pm0.09$ & Friel et al.~(\cite{friel05})\\
NGC6791  & 10.9         &8.12         &   Friel et al.~(\cite{friel02})          &   10-60  & 20,000       &15,gia    &synthesis& $+0.39\pm0.08$ & Carraro et al.~(\cite{carraro06}) \\ 

%
\hline
\hline
\end{tabular}
\label{tab_ocs}
\end{table*}
\end{landscape}

\setcounter{table}{2}
\begin{landscape}
\begin{table*}
\scriptsize
\label{table:grad}
\caption{ [Fe/H], [Ca/H], [Si/H], [Ti/H], [Ni/H], [Cr/H] gradients (dex~kpc$^{-1}$) in different radial
ranges.  The observed gradients are computed from the data discussed
in Sect.~\ref{sect_data} using a weighted linear fit. Model
gradients are approximated by linear fits. 
$^\dag$ For OCs with
$4~\mbox{Gyr}<\mbox{age}<11~\mbox{Gyr}$, the presence of the two metal rich OCs discussed in the text 
leads to a steep gradient ($-0.16$). Excluding them
we still obtain flatter gradients
($-0.13$), but still steeper than model predictions.  
$^{(a)}$ Andrievsky et al.~(\cite{andri04}) in the radial
ranges: 4.0--6.6~kpc, 6.6--10.6~kpc, 10.6--14.6~kpc.  $^{(b)}$ Lemasle et
al.~(\cite{lemasle07}) in the radial range 8--12~kpc.  $^{(c)}$ OCs from
Table~\ref{tab_ocs} and references therein, $^{(d)}$ in the radial range 7-12~kpc}      
\centering
\hspace{-5cm}
\begin{tabular}{lllllllll}        
\hline
\smallskip
          &                       &  model      &           &           & observations        &               &                &                 \\
\hline
gradient & range                 & present time & 4~Gyr ago & 11~Gyr ago & Cepheids             & OCs$^{(c)}$            &OCs$^{(c)}$             & OCs$^{(c)}$               \\ 
         &(kpc)                        &             &           &           & present time         & $\mbox{age}\leq 0.8$~Gyr   & $0.8~\mbox{Gyr}<\mbox{age}\leq 4~\mbox{Gyr}$ & $4~\mbox{Gyr}<\mbox{age}\leq 11~\mbox{Gyr}$  \\
\hline
\smallskip
$\Delta$[Fe/H]/$\Delta R$ &4--7   & $-0.134$    & $-0.162$  & $-0.176$  & $-0.128\pm0.029^{(a)}$  & --               & --               & --                \\
$\Delta$[Fe/H]/$\Delta R$ & 7--12 & $-0.075$    & $-0.071$  & $-0.076$  & $-0.044\pm 0.011^{(a)}$ & $-0.053\pm 0.029$  & $-0.094\pm 0.008$  & $-0.091\pm0.06$\\
                          &                     &           &              & $-0.051\pm 0.004^{(b)}$  &                  &                  &          \\
$\Delta$[Fe/H]/$\Delta R$ &12--22 & $-0.019$    & $-0.023$  & $-0.028$  & $+0.004\pm 0.011^{(a)}$  & --                 & $+0.005\pm 0.050$ &  $-0.001\pm 0.008$              \\
\hline
\smallskip
$\Delta$[Ca/H]/$\Delta R$ &4--7    & $-0.137$    & $-0.172$  & $-0.156$  &                       & --               & --               & --                \\
$\Delta$[Ca/H]/$\Delta R$ & 7--12  & $-0.073$    & $-0.067$  & $-0.076$ &                          &  --              & $0.02\pm0.03$    & $-0.13\pm0.03$        \\
$\Delta$[Ca/H]/$\Delta R$ &12--22  & $-0.018$   & $-0.024$  & $-0.029$  &                          &  --              & $-0.03\pm 0.09$  &  $0.009\pm 0.009$              \\
\hline
\smallskip
$\Delta$[Si/H]/$\Delta R$ &4--7    & $-0.124$    & $-0.153$  & $-0.126$  &                       & --               & --                    & --                \\
$\Delta$[Si/H]/$\Delta R$ & 7--12  & $-0.063$    & $-0.056$  & $-0.076$ &                          & $-0.05\pm 0.03$  & $-0.07\pm 0.10$       &  $-0.12\pm0.03$           \\
$\Delta$[Si/H]/$\Delta R$ &12--22  & $-0.017$   & $-0.024$  & $-0.029$  &                          &  --              &   --                  &  $0.003\pm 0.013$          \\
\hline
\smallskip
$\Delta$[Ti/H]/$\Delta R$ &4--7    & $-0.114$    & $-0.135$  & $-0.138$  &                       & --               & --                    & --                \\
$\Delta$[Ti/H]/$\Delta R$ & 7--12  & $-0.039$    & $-0.039$  & $-0.046$  &                         & $-0.08\pm 0.03$  & $-0.07\pm 0.06$        & $-0.16\pm0.04$            \\
$\Delta$[Ti/H]/$\Delta R$ &12--22  & $-0.007$    & $-0.009$  & $-0.018$  &                         &  --              & $-0.04\pm 0.02$       &  $0.02\pm 0.01$              \\
\hline
\smallskip
$\Delta$[Ni/H]/$\Delta R$ &4--7    & $-0.126$    & $-0.157$  & $-0.156$     &                       & --                & --                   & --                \\
$\Delta$[Ni/H]/$\Delta R$ & 7--12   & $-0.072$    & $-0.071$  & $-0.076$    &                         & --                & $-0.07\pm 0.06$      & $-0.13\pm0.03$             \\
$\Delta$[Ni/H]/$\Delta R$ &12--22  & $-0.020$   & $-0.024$  & $-0.029$      &                         & --                & --                   &  $0.00\pm 0.02$         \\
\hline
\smallskip
$\Delta$[Cr/H]/$\Delta R$ &4--7      & $-0.130$    & $-0.162$  & $-0.153$     &                       &    &        & --                \\
$\Delta$[Cr/H]/$\Delta R$ & 7--12    & $-0.072$    & $-0.068$  & $-0.075$&                         & $-0.08\pm 0.03$              & $0.04\pm 0.11$                   & $-0.18\pm0.03$             \\
$\Delta$[Cr/H]/$\Delta R$ &12--22    & $-0.020$   & $-0.024$  & $-0.029$   &                         & --                & --                    &  $-0.006\pm 0.03$              \\

\hline
\end{tabular}
\hspace{-5cm}
\label{tab_grad_iron} 
\end{table*}
\end{landscape}

\appendix

\section{The chemical evolution model of the Milky Way}
The model follows the evolution of the baryonic constituents (diffuse
gas, clouds, stars, and remnants) of the Galaxy (halo and disk) from
$t=0$, when all the mass of the Galaxy is in the form of diffuse gas
and it is located in the halo, until later times, when the mass fraction
in the various components is modified by several conversion processes.

The Galaxy is divided into $N$ coaxial cylindrical annuli with
inner and outer Galactocentric radii $R_i$ ($i=1,N$) and $R_{i+1}$,
respectively, mean radius $R_{i+1/2}\equiv (R_i+R_{i+1})/2$ and height
$h(R_{i+1/2})$.  Each annulus is divided into two zones: the {\em
halo} and the {\em disk}. These components are both made up of diffuse gas
$g$, clouds $c$, stars $s$ and stellar remnants $r$.  The {\em halo}
component is intended here quite generally as the material external to
the disk from which it is formed and accreted, i.e. the primordial
baryonic halo, and/or the material accreted from interactions with
small nearby dwarf galaxies or from the intergalactic medium during the
lifetime of the Galaxy.

The main processes considered in the model are: ({\em i}\/) conversion of diffuse
gas into clouds; ({\em ii}\/) formation of stars from clouds and ({\em
iii}\/) disruption of clouds by previous generations of massive stars;
({\em iv}\/) evolution of stars into remnants and return of a fraction
of their mass to the diffuse gas. 
We describe in the following the main equations of the model, using subscripts $H$ or $D$ to indicate the halo or the
disk, respectively.  For a spherical halo of radius $R_{N+1}\equiv
R_H$,
\begin{equation}
h(R_{i+1/2})=(R_H^2-R_{i+1/2}^2)^{1/2},
\end{equation}
and the halo volume in the $i$--th annulus is then  
\begin{equation}
V_{H,i}=\pi (R_{i+1}^2-R_i^2) h(R_{i+1/2}).
\end{equation}
If $R_D$ is the disk scale height (assumed independent of
radius), the volume of the disk in each annulus is
\begin{equation}
V_{D,i}=\pi (R_{i+1}^2-R_i^2) R_D.
\end{equation}
At time $t=0$ the Galaxy is constituted by diffuse gas
in the halo. At later times, the mass fraction in the various components is
modified by several conversion processes: diffuse gas is converted into
clouds, clouds collapse to form stars and are disrupted by massive
stars, stars evolve into remnants and return a fraction of their mass
to the diffuse gas. 
A disk of mass $M_D(t)$ is formed by continuous
infall from the halo of mass $M_H(t)$ at a rate
\begin{equation}
\frac{{\rm d}M_D}{{\rm d}t}=fM_H,
\end{equation}
where $f$ is a coefficient of the order of the inverse of the infall
time scale. Clouds condense out of diffuse gas at a rate $\mu$ and 
are disrupted by cloud-cloud
collisions at a rate $H^\prime$,
\begin{equation}
\frac{{\rm d}M_c}{{\rm d}t}=\mu M_g^{3/2}-H^\prime M_c^2,
\end{equation}
where $M_g(t)$ and $M_c(t)$ are the mass fractions of diffuse gas and
clouds, respectively.  Stars form in the halo and the disk by
cloud-cloud collisions at a rate $H$ and by the interactions of massive
stars with clouds at a rate $a$,
\begin{equation}
\frac{{\rm d}M_s}{{\rm d}t}=H M_c^2+aM_sM_c-DM_s,
\end{equation}
where $M_s(t)$ is the mass fraction in stars and $D$ is the stellar death rate.

Each annulus is then evolved independently (i.e. without radial mass flows)
keeping its total mass fixed from $t=0$ to $t_{\rm gal}=13.6$~Gyr, the age of the Galaxy. 
computing the fraction of mass in each component in the two zones, and
the chemical composition of the gas (assumed identical for diffuse gas
and clouds).  The rate coefficients of the model are all assumed to be
independent of time but functions of the Galactocentric radius $R$.
Their radial dependence  is discussed in
detail by Ferrini et al.~(\cite{ferrini94}), Moll\'a et
al.~(\cite{molla96}), and Moll\'a \& Diaz~(\cite{molla05}).  In
general, coefficients representing condensation processes (e.g.
the formation of clouds from diffuse gas), being proportional to the
inverse of the dynamical time, scale with the inverse square root of
the zone volume, whereas the coefficients of binary processes (e.g. 
the formation of stars by cloud-cloud collisions) scale as the
inverse of the zone volume; the coefficient of star formation induced
by stars is independent of radius.

\subsection{The infall}
\label{sec_infall}

The accretion rate of gas in the disk $\Sigma_g$, the infall law,  
is given in our model by 
\begin{equation}
\frac{\partial \Sigma_g(R)}{\partial t}=\frac{\Sigma_H(R)}{\tau} e^{-t/\tau} e^{-R/\lambda_D}, 
\label{eq_infall}
\end{equation}
which represents the infall on the disk resulting
from the collapse of the halo on a timescale $\tau$ with
radial lenghtscale $\lambda_D$.  
For $\tau$ and $\lambda_D$ we assume values typical for the solar
neighborhood, $\tau=8$~Gyr and $\lambda_D$=2~kpc (Sackett
\cite{sackett97}). The collapse timescale at the solar radius is
based on several observational constraints (e.g., G-dwarf metallicity
distribution, O/Fe vs. Fe/H relations, present-time infall rate), all
suggesting a time scale for the formation of the disk at the solar
circle of about 8~Gyr (see Moll\`a \& Diaz~\cite{molla05} and references
therein).

We have also explored the consequence of an additional infall rate constant in time and uniform in radius.
This is expressed by the following equation
\begin{equation}
\frac{\partial \Sigma_g(R)}{\partial t}=\frac{\Sigma_H(R)}{\tau} e^{-t/\tau} e^{-R/\lambda_D}
+\left(\frac{\partial \Sigma_g}{\partial t}\right)
\label{eq_infall1}
\end{equation}
where the first term is as in eq.~(\ref{eq_infall}) and the second
term is represents continuous accretion of matter.  This latter
component, when tuned to the appropriate value of
$\sim$1~$M_\odot$~yr$^{-1}$ integrated over 22~kpc radii, allows us to
reproduce better the observed abundance plateau in the outer Galaxy up
to $R_{\rm GC}\sim 22$~kpc, as shown in the paper.  On the other
hand, the addition of a constant infall produces a radial profile of
the SFR that is not consistent with current observations. In fact, in
this case, we would have at higher Galactocentric distance, a SFR
comparable to that at solar radius.  For this reason, even if the
outer flat gradient was well reproduced, we did not use this
parametrization.

\begin{figure} 
\psfig{figure=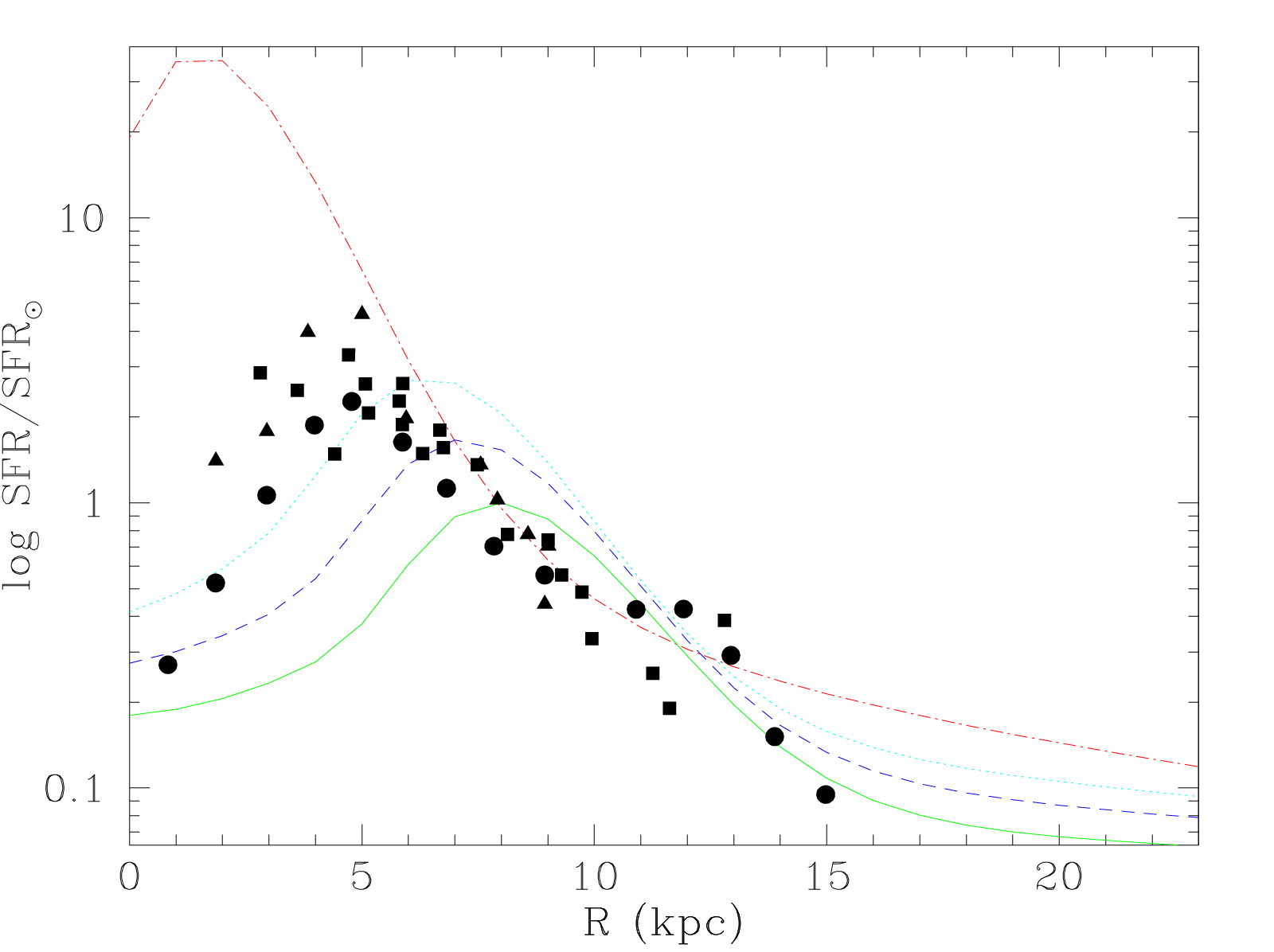, width=6cm, angle=-90} 
\caption{The radial profile of the SFR, normalized to its value at
R$_{\odot}$=8 kpc and at the present time obtained with 
an exponential infall law as in eq.~(\ref{eq_infall}).
The model curves are  
compared with the observations by Lacey \& Fall (1985)--filled
squares--, Gusten \& Mezger (1983)--filled triangles-- and Williams \&
McKee (1997)--filled circles. The model curves are at 1 (dotted-dashed line), 
5 (dotted line), 8 (dashed line) Gyr from the formation of the Galaxy, and the 
present time (continuous line).  }
\label{fig_sfr_infall} 
\end{figure}

\subsection{The parameters of the model}

The mass fractions in each component computed by the model in each annulus
are then converted into surface densities using as a normalization the
observed total baryonic surface density (i.e. the sum of the H{\sc
i}, H$_2$, stellar and remnant surface densities).  The observed
gas surface density is taken from Dame et al.~(\cite{dame93}), and
the stellar surface density, including remnants, from Boissier \&
Prantzos~(\cite{boissier99}), with an exponential scale-length of 2.5~kpc
(Ruphy et al.~\cite{ruphy96}) and a normalization value at the solar
radius of $43 \pm 5$~$M_\odot$~pc$^{-2}$ (Mera et al.~\cite{mera98}).
In the comparison with the data, we identify the diffuse gas and cloud
components with the H{\sc i} and the H$_2$ gas, respectively.

For our model of the Galaxy we adopt the ``best'' values of the various
coefficients (infall, cloud and star formation efficiencies) in the solar
neighborhood, scaling them at different Galactocentric radii, as discussed
by Ferrini et al.~(\cite{ferrini94}), Moll\`a et al.~(\cite{molla96}),
Moll\`a \& Diaz~(\cite{molla05}), Magrini et al.~(\cite{magrini07}).
The efficiencies of the processes regulating the conversion of diffuse gas
into clouds and clouds into stars are close to those found in the best
model for the Galaxy of Moll\`a \& Diaz~(\cite{molla05}), corresponding
to the efficiencies of their morphological type $N=4$ (see their Table~2).
In Table~\ref{Tab_model} we show the values of the coefficients
we have adopted for our best model. $R_D$ and $R_H$ are the scale
heights of the disk (cf. Sackett~\cite{sackett97}) and of the halo
(cf. Boissier \& Prantzos~\cite{boissier99}), respectively.  $\lambda_D$
is a typical disk scale length (cf. Ferrini et al. \cite{ferrini94},
Portinari \& Chiosi \cite{portinari99}). $\mu$ is the efficiency of
the cloud formation from diffuse gas, $H$ that of the star formation
from cloud collisions, and $H^\prime$ that of cloud dispersion.

\begin{table*}
\centering   
  \begin{tabular}{llllllll}
\hline
            & $h_D$         & $R_H$ & $\lambda_D$ &  $H (R_{\odot}$)       & $H^\prime (R_{\odot}$) 
  & $\mu (R_{\odot}$)                  & $f_0$                \\
            & (kpc)         & (kpc) & (kpc)       &(10$^{7}$~ yr)$^{-1}$& (10$^{7}$~yr)$^{-1}$&  (10$^{7}$~ yr) $^{-1}$ & (10$^{7}$~ yr)$^{-1}$         \\
\hline
MW           &  0.3    & 35  & 2.0   & 0.20    & 1.0   & 0.09    &  0.012\\
\hline
\end{tabular}
\caption{ Parameters of the GCE models for the MW. $R_D$ and $R_H$ are the scale heights of the disk and of the halo, respectively.
$\lambda_D$ is the scale length of the infall. 
$H$ (star formation efficiency from cloud collisions), $H^\prime$ (cloud dispersion) and $\mu$ (cloud formation), and $f$ (infall) 
are given at the solar radius ($R_{\odot}$).}
\label{Tab_model}  
\end{table*}

The chemical enrichment of the gas is modeled using the matrix formalism
developed by Talbot \& Arnett~(\cite{talbot73}).  The elements of
the restitution matrices $Q_{i,j}(M,Z)$ are defined as the fraction
of the mass of an element $j$ initially present in a star of mass $M$
and metallicity $Z$ that it is converted into an element $i$ and ejected.
Our GCE model takes into account two different metallicities, $Z=0.02$
and $Z=0.006$, and 22 stellar mass bins (21 for $Z=0.006$) from
$M_{\rm min}=0.8$~$M_\odot$ to $M_{\rm max}=100~$~$M_\odot$, for a
total of 43 restitution matrices.  For low- and intermediate-mass
stars ($M<8$~$M_{\odot}$) we use the yields by Gavil\'an et
al.~(\cite{gavilan05}) for both values of the metallicity.  For stars
in the mass range $8~M_\odot < M < 35~M_\odot$ we adopt the yields by
Chieffi \& Limongi~(\cite{chieffi04}) for $Z=0.006$ and $Z=0.02$.

We estimate the yields of stars in the mass range $35~M_\odot
<M<100~M_\odot$, which are not included in tables of Chieffi \&
Limongi~(\cite{chieffi04}) by linear extrapolation of the yields
in the mass range $8~M_\odot < M < 35~M_\odot$.  In the case
of Ti, we increased the yields by a factor of 2 to reproduce the
observations. The SNIa yields are taken from the model CDD1 by Iwamoto
et al.~(\cite{iwamoto99}).

Another fundamental ingredient of a GCE model is the initial mass
function (IMF). Several parameterizations of the IMF have been widely
used in the literature.  Since the magnitude and the slope of the modeled
chemical abundance gradients are related to the choice of the IMF, we
checked the consistency of several IMFs with the observed gradients.
With the adopted infall, efficiencies and set of chemical yields, we
have found the best agreement with the observed gradients using the IMF
by Kroupa et al.~(\cite{kroupa93}).

\subsection{Results of the model} 

A successful chemical evolution model of the Milky Way should
reproduce the observed radial profiles and the integrated quantities.
Fundamental radial profiles to be reproduced are: i) the gas radial
distribution (e.g. Dame~\cite{dame93}); ii) the stellar exponentially decreasing profile; iii) 
the strongly
decreasing SFR profile; iv) the metallicity gradients of various
elements.  The integrated quantities are instead the average infall
rate, the average SFR, and the supernovae rates of both SNII and SNIa.

\begin{figure} 
\psfig{figure=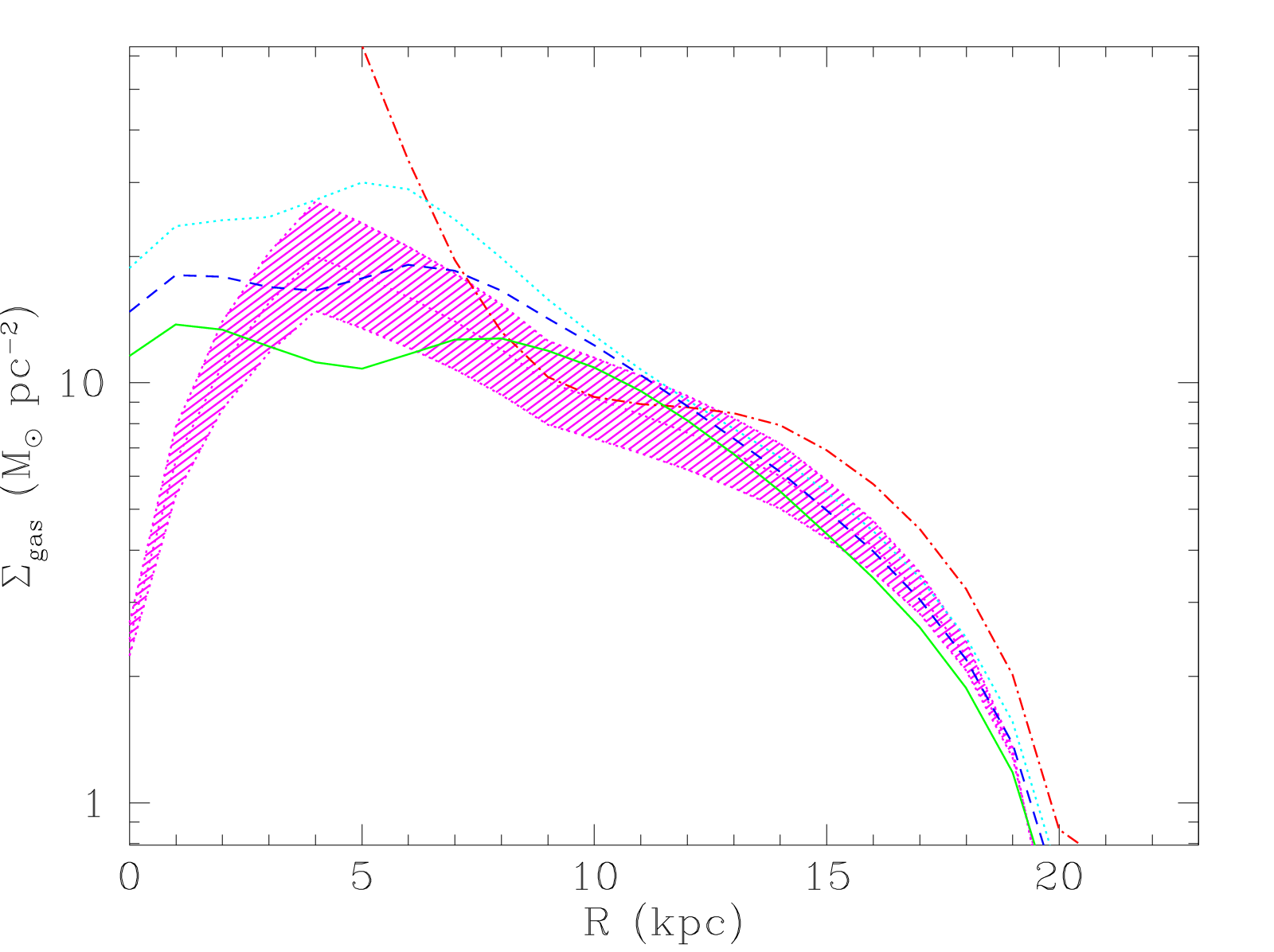, width=6cm, angle=-90} 
\psfig{figure=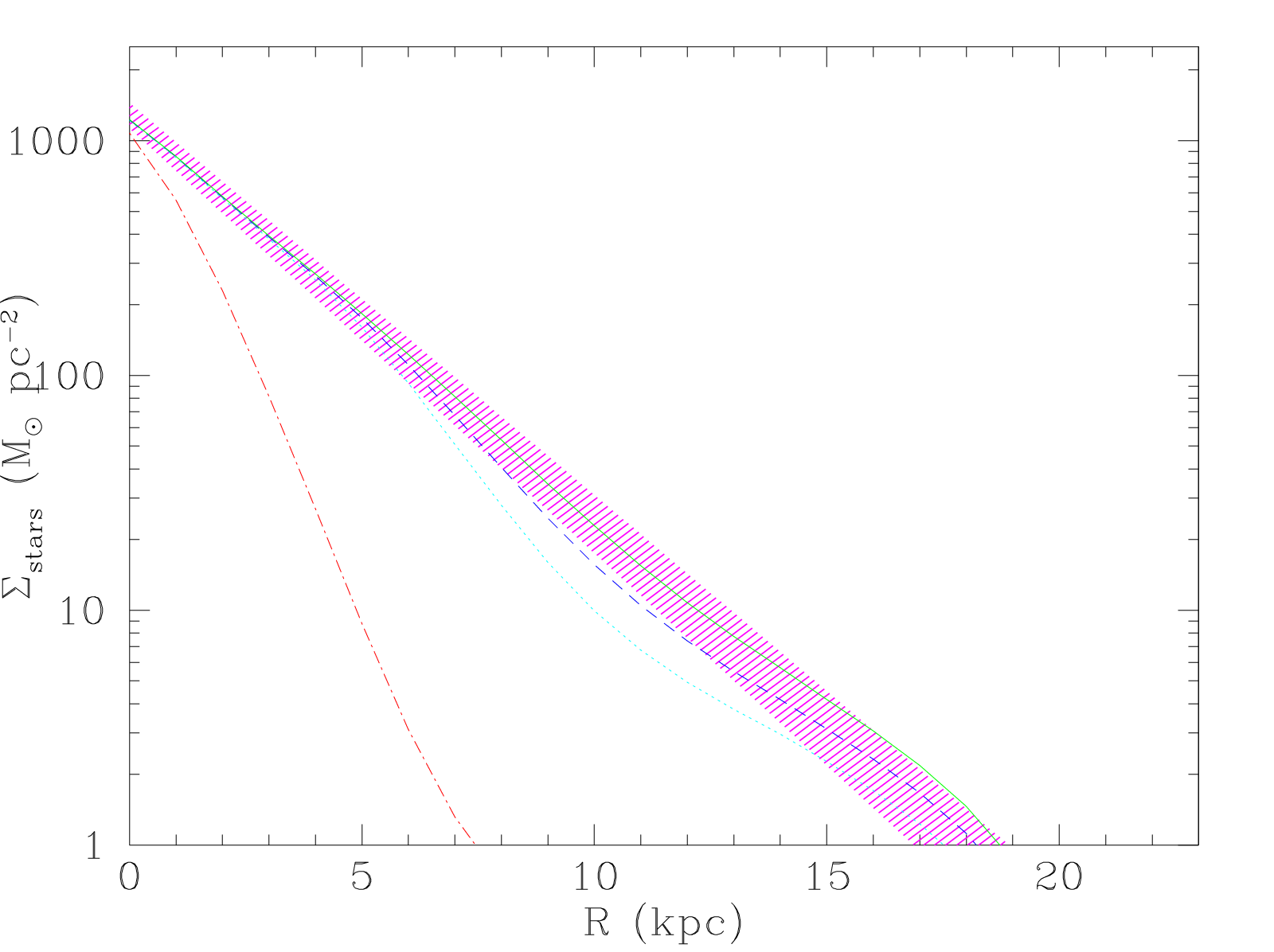, width=6cm, angle=-90} 
\caption{Results of the chemical evolution model for the Galactic 
disk and comparison to observations  
a) Total gas profile, model at 1, 5, 8 Gyr, and present time from the formation of the Galaxy, observations 
by Dame (\cite{dame93}); b) Stellar
profile, model as in a), and observations by Mera et al.~(\cite{mera98}), within two exponential disks
of scale-lengths 2.2 and 2.6 kpc, respectively and normalized to the
current local star surface density; line types for the model are as in Fig.~\ref{fig_sfr_infall}.}
\label{fig_model} 
\end{figure}

\begin{figure} 
\psfig{figure=0634fig16.eps, width=6cm, angle=-90} 
\psfig{figure=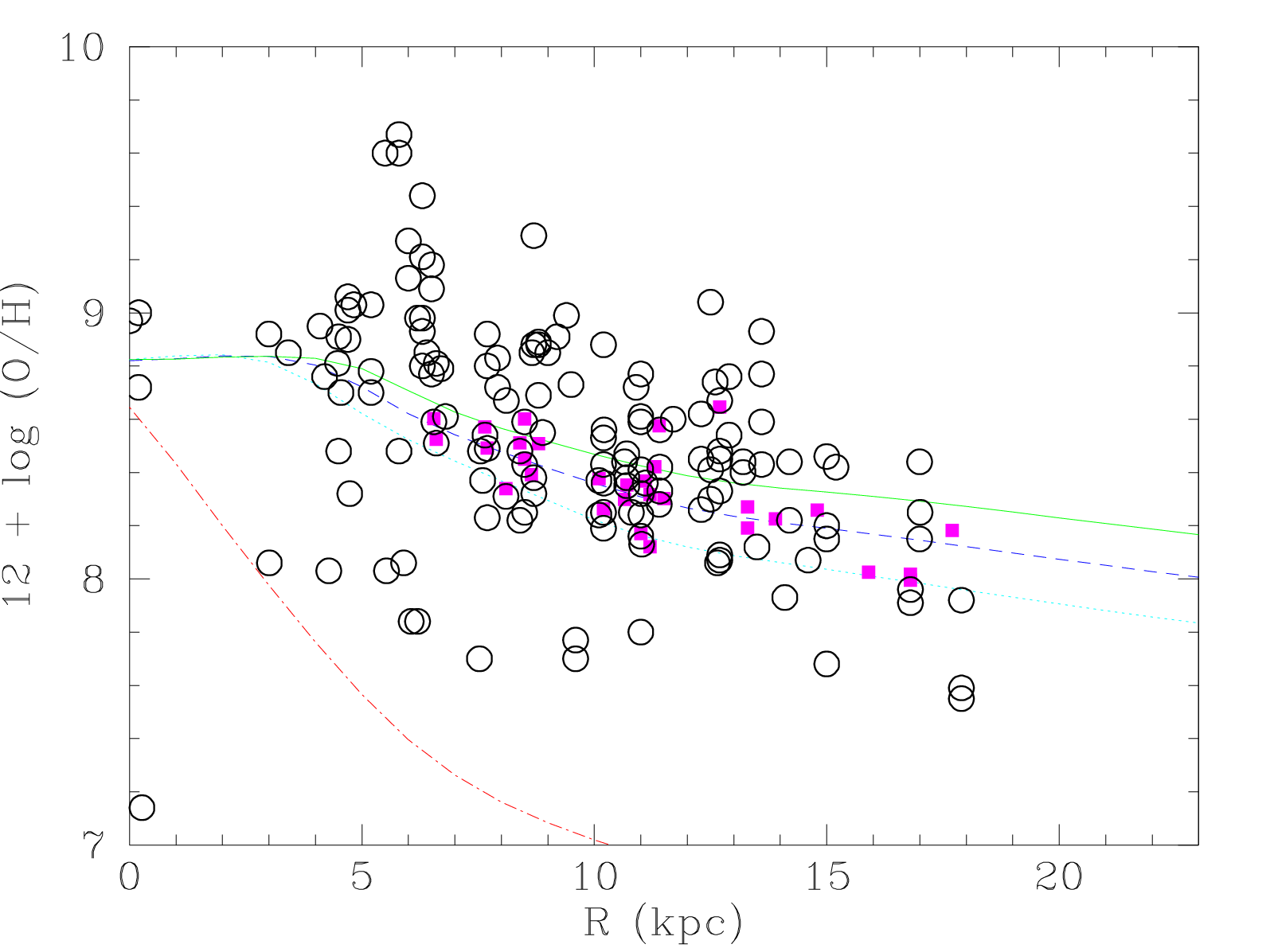, width=6cm, angle=-90} 
\caption{Results of the chemical evolution model for the Milky Way
disk, model at different times as in Fig.\ref{fig_model}. 
a) Oxygen abundance gradient  compared with the 
HII regions gradients by Deharveng et al.~(\cite{deharveng00}, filled squares) and Rudolph et al.~(\cite{rudolph06}), empty circles).
b) Radial profile of the infall of gas into the disk  Line types for the model are as in Fig.~\ref{fig_sfr_infall}.}
\label{fig_model1} 
\end{figure}

In Figures~\ref{fig_model} and \ref{fig_model1}  we present the comparison of the results of our model with 
the previously described radial profiles. 
Note that the model describes only the evolution of the disk,  thus the inner regions hosting the bulge are 
not expected to be reproduced  (R$\lesssim$5~kpc).
The radial profiles of the total gas and of the stars at the present time
are well reproduced, as is the radial profile of O/H as drawn by
\hii\ regions. Note however the large dispersion in the observations.
The general trend of the SFR is reproduced by the model, even if the
observed maximum is at $\sim$6-7~kpc, while the model maximum is located  at $\sim$9-10~kpc

Our ``best'' model is also able to reproduce the observational
integrated quantities for the Galaxy, as: ({\em i}\/) the present-time
infall rate is 3.5~$M_\odot$~yr$^{-1}$ integrated over 22~kpc radii,
consistent with the recent estimate $\sim 5$~$M_\odot$~yr$^{-1}$ by de
Boer~(\cite{deboer04}), and favoured by recent models of disk galaxy
evolution (e.g., Naab \& Ostriker~\cite{naab06}); ({\em ii}\/) the
predicted current star formation rate is $\sim 3$~$M_\odot$~yr$^{-1}$
integrated over 22~kpc radii, to be compared with the observed value
e.g. by Pagel (\cite{pagel97}) of 2-6~$M_\odot$~yr$^{-1}$; ({\em
iii}\/) the current supernova rates obtained from the model are
1.6~SN/century for SNIa and 1.4~SN/century for SNII, of the same order as
the values for Sb galaxies $\sim 1.8$~SN/century and $\sim 2$~SN/century,
respectively (Mannucci et al.~\cite{mannucci05}).

\begin{acknowledgements}
We thank the anonymous referee for useful comments which have improved the paper.
This research has made use of the WEBDA database, operated at the
Institute for Astronomy of the University of Vienna.  LM is supported
by an INAF post-doctoral grant 2005.  DG acknowledges support from the
EC Research Training Network MRTN-CT-2006-035890 ``Constellation''.
\end{acknowledgements}

{}

\end{document}